 \documentclass[final,5p,times,twocolumn,numbers]{elsarticle}

\usepackage{amssymb}
\usepackage{lipsum}
\usepackage{xcolor}
\usepackage{amsmath}
\usepackage{needspace}
\usepackage{cuted}
\usepackage{mathrsfs}
\usepackage{subcaption}

\journal{Physics Letters B}

\begin{document}

\begin{frontmatter}



\title{Impact of Rastall gravity on hydrostatic mass of galaxy clusters}

\def\Journal#1#2#3#4{{ #1} {\bf #2}, #3 (#4) }
\def\RPP{{Rep. Prog. Phys}}
\def\PRC{{Phys. Rev. C}}
\def\PRD{{Phys. Rev. D}}
\def\ZPA{{Z. Phys. A}}
\def\NPA{{Nucl. Phys. A}} 
\def\JPG{{J. Phys. G }}
\def\PRL{{Phys. Rev. Lett}}
\def\PR{{Phys. Rep.}}
\def\PREV{{Phys. Rev.}}
\def\PRX{{Phys. Rev. X}}
\def\PLB{{Phys. Lett. B}}
\def\AP{{Ann. Phys (N.Y.)}}
\def\EPJA{{Eur. Phys. J. A}}
\def\NP{{Nucl. Phys}}  
\def\RMP{{Rev. Mod. Phys}}
\def\IJMPE{{Int. J. Mod. Phys. E}}
\def\AJ{{Astrophys. J}}
\def\AJL{{Astrophys. J. Lett}}
\def\AA{{Astron. Astrophys}}
\def\ARAA{{Annu. Rev. Astron. Astrophys}}
\def\MPLA{{Mod. Phys. Lett. A}}
\def\ARNPS{{Annu. Rev. Nuc. Part. Sci}}
\def\LRR{{Living. Rev. Relativity}}
\def\CARAA{{Class. Ann. Rev. Astron. Astrophys.}}
\def\EPJC{{Eur. Phys. J. C}}
\def\cqg{{Class. Quantum Grav.}}
\def\mon{{Mon. Not. R. Astron. Soc.}}
\def\AJSS{{Astrophys. J. Suppl. Ser.}}

 \author[label1,label2]{M. Lawrence Pattersons}
 \ead{m.pattersons@proton.me} 
 \author[label3,label4,label5,label6]{Feri Apryandi}
 \ead{feri.apryandi@upi.edu}
 \author[label1,label2]{Freddy P. Zen}
 \ead{fpzen@fi.itb.ac.id}
 \affiliation[label1]{organization={Theoretical High Energy Physics Group, Department of Physics, Institut Teknologi Bandung},
             addressline={Jl. Ganesha 10},
             city={Bandung},
             postcode={40132},
             country={Indonesia}}
\affiliation[label2]{organization={Indonesia Center for Theoretical and Mathematical Physics (ICTMP), Institut Teknologi Bandung},
             addressline={Jl. Ganesha 10},
             city={Bandung},
             postcode={40132},
             country={Indonesia}}
 \affiliation[label3]{organization={Physics Study Program, Faculty of Mathematics and Natural Science Education, Universitas Pendidikan Indonesia},
             city={Bandung},
             country={Indonesia}}
\affiliation[label4]{organization={Research Group on Environmental Exploration, Mitigation, Education, and Defense of Earth \& Outer Space, Universitas~Pendidikan~Indonesia},
             city={Bandung},
             country={Indonesia}}
\affiliation[label5]{organization={Learning Analytics and Digital Assessment Research Group, Universitas Pendidikan Indonesia},
             city={Bandung},
             country={Indonesia}}
\affiliation[label6]{organization={Center of Excellence Astronomical Data Science and Light Pollution, Universitas Pendidikan Indonesia},
             city={Bandung},
             country={Indonesia}}


\begin{abstract}
Galaxy clusters are the largest virialized structures in the Universe and are predominantly dominated by dark matter. The hydrostatic mass and the mass obtained from gravitational lensing measurements generally differ, a discrepancy known as the hydrostatic mass bias. \textcolor{black}{In this work, we derive the hydrostatic mass of galaxy clusters within the framework of Rastall gravity. We consider two scenarios: (i) the absence of dark matter and (ii) the presence of dark matter. In both cases, we constrain the Rastall parameter in the cluster-scale using  observational data.} In the first scenario, Rastall gravity effectively reduces the hydrostatic mass, bringing it closer to the observed baryonic mass. The best linear fit yields a slope $\mathbf{M}=1.07\pm0.11$, indicating a near one-to-one correspondence between the two masses. In the second scenario, Rastall gravity helps to alleviate the hydrostatic mass bias. The linear fit between the Rastall hydrostatic mass and the observed lensing mass results in a best-fit slope \textcolor{black}{$\mathbf{M}=0.99\pm0.26$}, which is very close to unity. \textcolor{black}{We also calculate the goodness-of-fit for every fit. The statistical evaluations indicate that Rastall gravity provides a viable phenomenological framework that can improve certain aspects of the mass discrepancy problem at the level of scaling relations. However, it does not universally outperform other modified gravity model, when evaluated using standard goodness-of-fit criteria.}
\end{abstract}




\end{frontmatter}




\section{Introduction}
\label{intro}
Galaxy clusters are the largest virialized astronomical objects in the universe~\cite{Holanda2015}. The formation of galaxy clusters is related to the collapse of the largest gravitationally bound overdensities in the initial density field and is accompanied by the most energetic phenomena since the Big Bang and by the complex interplay between gravity-induced dynamics of collapse and baryonic processes associated with galaxy formation~\cite{Kravtsov2012}. The population of galaxies in clusters has evolved rapidly over the last 5 Gyr~\cite{Mercurio2021}. Galaxy clusters host a hot and diffuse intracluster medium, characterized by typical temperatures $\sim 7\times10^7$ K and electron density $\sim10^{-3}$ cm$^{-3}$~\cite{Apryandi2025,Fabian1992}.

The mass of galaxy clusters can be determined using three main methods: velocity dispersion measurements, X-ray or Sunyaev-Zel’dovich (SZ) observations under the assumption of hydrostatic equilibrium, and gravitational lensing. Except for gravitational lensing, these methods estimate the cluster mass within the framework of the Newtonian approximation~\cite{Apryandi2025,Desai2020}. The reviews of the aforementioned methods are available in Refs.~\cite{Kravtsov2012,Sarazin1986,Bartelmann1995,Voit2005,Allen2011,Ettori2013,Hoekstra2013,Munari2013,Vikhlinin2014,Li2023}. \textcolor{black}{The hydrostatic mass approach has been commonly used in the studies of galaxy clusters~\cite{Ettori2019,Pearce2020,Scheck2023,Nelson2014,Turner2025,Braspenning2025}}. \textcolor{black}{Note that there is another method to calculate mass of galaxy clusters, i.e. richness proxy. Here, richness is defined as the number of bright cluster galaxies~\cite{Andreon2015}. The details of the richness proxy can be found in Refs.~\cite{Andreon2015,Chiu2019,Andreon2016,Andreon2010,Andreon2012}}

It should be noted that generally, the lensing mass of galaxy clusters is systematically higher than the hydrostatic mass. This discrepancy is commonly referred to as the hydrostatic mass bias~\cite{Desai2022}. In terms of mass composition, galaxy clusters are dominated by dark matter (DM)~\cite{Cerini2025}, which can account for approximately $80-90\%$ of their total mass~\cite{Stepanov2025}. Based on computational simulations, galaxy clusters are identified as the largest gravitationally bound DM halos in the large-scale structure of the Universe~\cite{Premadi2021}.

Bambi~\cite{Bambi2007} conducted a detailed study on the mass determination of galaxy clusters. Assuming that dark energy (DE) behaves as a cosmological constant $\Lambda$ on relatively small scales, he investigated the effects of a nonzero cosmological constant in the Newtonian limit. As a result, he derived a general relativity (GR) correction to the mass of galaxy clusters, leading to an effective mass for galaxy clusters. Desai~\cite{Desai2022} computed the masses of several galaxy clusters using Bambi’s formulation and showed that, within the considered samples, the impact of general relativistic corrections on the hydrostatic masses of galaxy clusters is negligible.

Gupta and Desai~\cite{Desai2020} also evaluated the general relativistic correction to the mass of galaxy clusters using the Tolman-Oppenheimer-Volkoff (TOV) equation, which represents hydrostatic equilibrium in general relativity. They similarly found that the GR correction is negligible, indicating that the standard Newtonian approach provides an adequate description.

On the other hand, the domination of the DE and DM in the our dark Universe has become a motivation to construct alternative theories beyond GR~\cite{Clifton2012}. Rahvar and Mashhoon~\cite{Mashhoon2014} calculated the masses of galaxy clusters within the framework of nonlocal gravity. They performed the linear fit between the nonlocal and baryonic mass estimates. The ideal scenario, in which nonlocal gravity fully accounts for the missing mass without involving DM within the scale of clusters, corresponds to $\mathbf{M}= 1$. By comparing the resulting nonlocal masses with the corresponding baryonic masses for a sample of ten galaxy clusters, the best-fit slope was found to be $\mathbf{M} = 0.84 \pm 0.04$. Using a similar approach but within the framework of Scalar–Tensor–Vector Gravity (STVG), Moffat and Rahvar~\cite{Moffat2014} obtained a value of $\mathbf{M}$ that is closer to unity than that reported by Rahvar and Mashhoon. They conclude that STVG is compatible with observational data on megaparsec scales without invoking DM.

Another attempt to utilize modified theories of gravity to calculate the masses of galaxy clusters was carried out by the authors in Ref.~\cite{Apryandi2025}. They employed Eddington-inspired Born–Infeld (EiBI) theory, beyond Horndeski gravity (BHG), and Newtonian gravity modified by the generalized uncertainty principle (GUP). The use of GUP in the calculation of the mass of galaxy clusters can also be found in Ref.~\cite{Apryandi2021}. Among these theories, EiBI theory yields the best linear fit with $\mathbf{M} = 0.126 \pm 0.086$. This value is significantly lower than that reported by Rahvar and Mashhoon and is also much farther from unity. Consequently, EiBI theory, BHG, and Newtonian gravity modified by the GUP cannot satisfactorily account for the masses of galaxy clusters without involving DM.

Another modified theory of gravity that is worth considering is Rastall gravity, which was introduced by Rastall~\cite{Rastall1972} in 1972. It allows for a nonconserved energy-momentum tensor (EMT), characterized by a nonvanishing covariant divergence of the EMT, which Rastall proposed as a generalization of GR. The violation of EMT conservation can be phenomenologically caused by quantum effects in classical context~\cite{Oliveira2015,Sakti2022,Pattersons2025a}. In contrast, the covariant divergence of the EMT identically vanishes in GR.

From a different perspective, Visser~\cite{Visser2018} claimed that Rastall gravity is equivalent to GR. He argued that, in the matter sector, Rastall’s EMT can be interpreted as an artificially isolated component of a physically conserved EMT. Consequently, the resulting theory does not introduce new gravitational dynamics and is therefore equivalent to GR.

Darabi et al.~\cite{Darabi2018} refuted Visser’s viewpoint. They emphasized that Rastall did not introduce new forms of the EMT in his original work. Instead, Rastall proposed a modified relation between the EMT and spacetime geometry, motivated by a possible unknown mutual interaction between them. They further stressed that violations of energy-momentum conservation have also been considered in theoretical frameworks distinct from Rastall gravity, for instance in attempts to model DE.

Despite this debate, Rastall gravity has been widely explored in the literature on modified theories of gravity. It has been applied to various astrophysical objects, including neutron stars~\cite{Oliveira2015,Pattersons2025a,Pattersons2025b,daSilva2021,Xi2020}, quark stars~\cite{Rizaldy2019,Tangphati2024,Banerjee2025,Salako2021,Banerjee2023}, wormholes~\cite{Bhar2024,Malik2023,Halder2019,Moradpour2017,Heydarzade2023,Bronnikov2021}, black holes~\cite{Sakti2022,Sakti2020,Nashed2022,Prihadi2020,Spallucci2018,Heydarzade2017a,Zou2020,Kumar2018,Heydarzade2017b}, and gravastars~\cite{Majeed2022}, as well as to cosmology in the context of an accelerated expanding Universe~\cite{Capone2010}. On galactic scales, Rastall gravity has been investigated by Li et al.~\cite{Li2019}, who modeled both stellar matter and DM as perfect fluids. Tang et al.~\cite{Tang2020} constrained the Rastall parameter using rotation curves of low surface brightness galaxies.

The aims of this work are twofold. First, we construct the formalism for the hydrostatic mass of galaxy clusters within the framework of Rastall gravity. \textcolor{black}{Second, we determine the best-fit values of in two scenarios: (i) the absence of DM, and (ii) the presence of DM; in order to constrain the Rastall parameter at cluster scales in both cases. In the first scenario, the Rastall-modified hydrostatic mass is compared with the observed baryonic mass of galaxy clusters. In the second scenario, the Rastall-modified hydrostatic mass is compared with the lensing mass data to examine whether Rastall gravity can alleviate the hydrostatic mass bias. We note that alleviating the discrepancy between hydrostatic and lensing mass estimates remains an open issue, as it was not resolved in Ref.~\cite{Desai2022}.} We find that the Rastall parameter has a nontrivial influence on the hydrostatic mass of galaxy clusters, and that the resulting best-fit values of $\mathbf{M}$ are close to unity, with $\mathbf{M}=1$ lying within the associated uncertainty range.

The structure of this paper is as follows. Section~\ref{formalism} presents the mathematical formalisms employed in this work, including Rastall gravity, the standard hydrostatic formulation for determining the mass of galaxy clusters, and the hydrostatic mass formulation within the framework of Rastall gravity. Section~\ref{numerical} presents the numerical results \textcolor{black}{and statistical evaluations on the results}, followed by a discussion of the corresponding implications. Finally, our conclusions are summarized in Section~\ref{conclusion}.

\section{Formalisms}
\label{formalism}
To ensure completeness, we present the formalism of Rastall gravity in Subsection~\ref{Rastall}, the standard hydrostatic formulation for determining the mass of galaxy clusters in Subsection~\ref{standard}, and the formulation of the hydrostatic mass of galaxy clusters within the framework of Rastall gravity derived in Subsection~\ref{massrastall}.

\subsection{Rastall gravity}
\label{Rastall}
First, in this Subsection, we use $G=c=1$, where $G$ is the universal gravitational constant, while $c$ denotes the speed of light. Also, we emphasize that, unlike most modified theories of gravity, which are formulated starting from an action principle, Rastall gravity is instead based on a modification of the divergence of the EMT. In GR, it is widely known that the divergence of EMT $T^{\mu\nu}$ reads
\begin{equation}
    \nabla_\mu T^{\mu\nu}=0,
\end{equation}
which ensures the EMT conservation. However, Rastall~\cite{Rastall1972} proposed a modification to the divergence of the EMT, given by
\begin{equation}
\nabla_\mu T^{\mu\nu} = \lambda \nabla^\nu \mathcal{R}, \label{rastalllaw}
\end{equation}
where $\lambda$ denotes the Rastall free parameter and $\mathcal{R}$ is the Ricci scalar. \textcolor{black}{As a consequence of Eq.~(\ref{rastalllaw}), the EMT is no longer conservative.} It is evident that in the limit $\lambda = 0$, Rastall gravity is reduced to GR. Thus, $\lambda$ represents the deviation of Rastall gravity from GR.

With Eq.~(\ref{rastalllaw}) in our hands, we can obtain the Rastall field equation, which writes~\cite{Pattersons2025a}
\begin{equation}
    G^\mu_\nu+\kappa\lambda \mathcal{R}\delta^\mu_\nu=\kappa T^\mu_\nu.\label{rastallfieldeq}
\end{equation}
Here, $G^\mu_\nu=\mathcal{R}^\mu_\nu-\frac{1}{2}\mathcal{R}\delta^\mu_\nu$ denotes the Einstein tensor, $\mathcal{R}^\mu_\nu$ denotes the Ricci tensor, $\delta^\mu_\nu$ is the Kronecker delta, and $\kappa$ is the constant of the field equation. For 4-dimensional spacetime, Eq.~(\ref{rastallfieldeq}) gives
\begin{equation}
    \mathcal{R}=\frac{\kappa T}{4\kappa\lambda-1},\label{riccirastall}
\end{equation}
where $T$ denotes the trace of EMT. \textcolor{black}{Consequently, the Rastall non-conservation law can be written as}
\begin{eqnarray}
    \textcolor{black}{\nabla_\mu T^{\mu\nu}=\frac{\kappa\lambda}{4\kappa\lambda-1}\nabla^\nu T}.\label{conservt}
\end{eqnarray}
Furthermore, Rastall field equation now becomes
\begin{equation}
    G^\mu_\nu=\kappa\mathcal{T}^\mu_\nu,
\end{equation}
where $\mathcal{T}^\mu_\nu$ is the effective modified EMT that satisfies \textcolor{black}{the definition}
\begin{equation}
    \mathcal{T}^\mu_\nu\textcolor{black}{\equiv}T^\mu_\nu-\delta^\mu_\nu\left(\frac{\kappa\lambda}{4\kappa\lambda-1}T\right).
\end{equation}

Before proceeding, it is worth pointing out that Refs.~\cite{Oliveira2015,Sakti2022,Prihadi2020} adopted $\kappa = 8\pi$ in Rastall gravity, as in GR. Nevertheless, the constant in Rastall gravity is intrinsically modified and differs from its GR counterpart. The authors of Ref.~\cite{daSilva2021} derived the constant $\kappa$ in Rastall gravity by applying the theory in the weak-field limit. They obtained
\begin{equation}
\kappa = \frac{8\pi}{2\gamma + 1},\label{kapparastall}
\end{equation}
where the parameter $\gamma$ is related to the Rastall parameter $\lambda$ through
\begin{equation}
\lambda = \frac{\gamma}{\kappa (4\gamma - 1)}.\label{lambdarastall}
\end{equation}
From Eqs.~(\ref{riccirastall}), (\ref{kapparastall}), and (\ref{lambdarastall}), we have an insight that $\kappa\lambda=1/4$, $\gamma=-1/2$, and $\kappa\gamma=1/4$ are forbidden, \textcolor{black}{since the denominator will be zero}. \textcolor{black}{In this work, the values of $\gamma$, $\kappa$, and $\lambda$ are shown in Table~\ref{valuekappa}. It is worth noting that those values are the parameters that result in the best-fits.}

\begin{table}[h]
\centering
\caption{\textcolor{black}{Values of $\gamma$, $\kappa$, and $\lambda$ used in this work.}}
\begin{tabular}{ccc}
\hline
\textcolor{black}{$\gamma$} & \textcolor{black}{$\kappa$} & \textcolor{black}{$\lambda$} \\
\hline
\textcolor{black}{$4.00$} & \textcolor{black}{$0.89\pi$} & \textcolor{black}{$9.55\times10^{-2}$} \\
\textcolor{black}{$4.95$} & \textcolor{black}{$0.73\pi$} & \textcolor{black}{$1.14\times10^{-1}$} \\
\textcolor{black}{$-3.50\times10^{-2}$} & \textcolor{black}{$8.60\pi$} & \textcolor{black}{$1.14\times10^{-3}$} \\
\textcolor{black}{$-4.10\times10^{-2}$} & \textcolor{black}{$8.71\pi$} & \textcolor{black}{$1.29\times10^{-3}$} \\
\hline
\end{tabular}
 \label{valuekappa}
\end{table}

\textcolor{black}{Note that the original Rastall's theory~\cite{Rastall1972} was derived by re-questioning the conservation of the EMT, rather than being derived from the action. Despite this fact, Ref.~\cite{demoraes2019} successfully derived the Rastall-type field equation from $f(\mathcal{R},T)$ gravity. In $f(\mathcal{R},T)$ gravity, the action writes~\cite{Yusmantoro2025}}
\begin{eqnarray}
    \textcolor{black}{S=\frac{1}{2\kappa}\int d^4 x \sqrt{-g} \bigg(f(\mathcal{R},T)+\mathscr{L}_m\bigg).}
\end{eqnarray}
\textcolor{black}{Here, $g$ denotes the determinant of the metric tensor, and $\mathscr{L}_m$ is the Lagrangian of matter.}

\textcolor{black}{In Ref.~\cite{demoraes2019}, a specific function $f(\mathcal{R},T)$ is chosen, i.e.}
\begin{eqnarray}
    \textcolor{black}{f(\mathcal{R},T)=\mathcal{R}+\kappa \varsigma T}.\label{frt}
\end{eqnarray}
\textcolor{black}{Here, $\varsigma$ is the free parameter of the $f(\mathcal{R},T)$ gravity theory. By deriving the action containing the function shown in Eq.~(\ref{frt}), one obtains the Rastall-type field equation, i.e.}
\begin{eqnarray}
    \textcolor{black}{G_{\mu\nu}=\frac{\kappa}{1-\varsigma}\bigg[(1-\varsigma)T_{\mu\nu}+\varsigma\bigg(\frac{g_{\mu\nu}}{2}T-g^{\alpha\beta}\frac{\delta T_{\alpha\beta}}{\delta g^{\mu\nu}}\bigg)\bigg].}
\end{eqnarray}
\textcolor{black}{For the static spherically symmetric perfect fluid EMT $T_{\mu\nu}=(\rho+p)u_\mu u_\nu+p g_{\mu\nu}$, where $\rho$, $p$, and $u_\mu$ denote energy density distribution of matter, pressure, and 4-velocity, respectively; we can easily find that}
\begin{eqnarray}
    \textcolor{black}{g^{\alpha\beta}\frac{\delta T_{\alpha\beta}}{\delta g^{\mu\nu}}=-2(\rho+p)u_\mu u_\nu-p g_{\mu\nu}.\label{tensortambahan}}
\end{eqnarray}
\textcolor{black}{By defining $\mathcal{T}^* =g^{\mu\nu}\bigg(g_{\mu\nu }T-2g^{\alpha\beta}\frac{\delta T_{\alpha\beta}}{\delta g^{\mu\nu}}\bigg)$, one obtains}
\begin{eqnarray}
    \textcolor{black}{\nabla_\mu T^{\mu\nu}=\frac{\varsigma}{2(\varsigma-1)}\nabla^\nu \mathcal{T}^*,}
\end{eqnarray}
\textcolor{black}{which is similar to Eq.~(\ref{conservt}) and exhibits the same interpretation, i.e., the non-conservation of EMT.}

\textcolor{black}{Within the weak-field regime, using Eq.~(\ref{tensortambahan}) and assuming a static, spherically symmetric perfect fluid with $\rho \gg p$ and $r \gg m$, we obtain $\mathcal{T}^*=-5\rho$. By choosing}
\begin{align}
\textcolor{black}{\varsigma=-\frac{2}{5\bigg[1-\frac{2\kappa\lambda}{5(4\kappa\lambda-1)}\bigg]}\frac{\kappa\lambda}{4\kappa\lambda-1},}
\end{align}
\textcolor{black}{and recalling that $T=-\rho$ in the weak-field regime, we obtain}
\begin{align}
    \textcolor{black}{\nabla_\mu T^{\mu\nu}}\textcolor{black}{=}&\textcolor{black}{-\frac{\kappa\lambda}{4\kappa\lambda-1}\nabla^\nu \rho,}\nonumber\\
    \textcolor{black}{=}&\textcolor{black}{\frac{\kappa\lambda}{4\kappa\lambda-1} \nabla^\nu T,}
\end{align}
\textcolor{black}{which is nothing but the Rastall non-conservation law shown by Eq.~(\ref{conservt}). At this stage, we have shown that, for a static spherically symmetric perfect fluid, Rastall gravity in the weak-field limit can be effectively derived from $f(\mathcal{R},T)$ gravity.}

\subsection{Standard hydrostatic mass calculation}
\label{standard}
We start from the hydrostatic equilibrium equation, given by~\cite{Schneider}
\begin{equation}
    \frac{dp}{dr}=-\rho\frac{Gm(r)}{r^2},\label{standardhydrostatic}
\end{equation}
where $m(r)$ is the mass of the clusters enclosed within a radius $r$. Here, the equation of state (EOS) of the galaxy clusters is assumed to be satisfied by the EOS of the ideal gas. Such EOS is employed by Ref.~\cite{Apryandi2025,Desai2020}. For the ideal gases, we have~\cite{Apryandi2025}
\begin{equation}
    \frac{dp}{dr}=\frac{k_B}{m_p\mu}\left(\tau\frac{d\rho}{dr}+\rho\frac{d\tau}{dr}\right).\label{idealgas}
\end{equation}
Here, $k_B$ denotes the Boltzmann constant, $m_p$ is the mass of proton, $\mu$ denotes the mean molecular weight of the cluster, $\tau$ is the absolute temperature.

By combining Eq.~(\ref{standardhydrostatic}) and Eq.~(\ref{idealgas}), one obtains
\begin{eqnarray}
    -\rho\frac{Gm(r)}{r^2}&=&\frac{k_B}{m_p\mu}\left(\tau\frac{d\rho}{dr}+\rho\frac{d\tau}{dr}\right)\nonumber\\
m(r)&=&-\frac{k_B r}{\mu m_p G}\tau\left(\frac{d\ln\rho}{d\ln r}+\frac{d\ln\tau}{d\ln r}\right).\label{recover}
\end{eqnarray}

By substituting the values of the constants $k_B$, $\mu$, $m_p$, and $G$, we have the expression of the hydrostatic mass of galaxy clusters, i.e.
\begin{eqnarray}
m(r)=-3.7\times10^{-4}\tau\:r\left(\frac{d\ln\rho}{d\ln r}+\frac{d\ln\tau}{d\ln r}\right)\mathcal{M}_\odot.\label{masstandardhyd}
\end{eqnarray}
Here, $\mathcal{M}_\odot=10^{14}\:M_\odot$.

\textcolor{black}{Now we proceed to the profiles related to the galaxy clusters. According to Ref.~\cite{Mashhoon2014}, the temperature of the intracluster gas is typically of the order of keV. At such temperatures, the hot plasma predominantly emits X-rays through thermal bremsstrahlung (free--free emission). In addition, line emission from ionized heavy elements also contributes to the overall radiation. The total emitted radiation is approximately proportional to the product of the electron number density $n_e$ and proton number density $n_p$.} Following Ref.~\cite{Mashhoon2014}, the hot gas density of galaxy clusters is given by
\begin{eqnarray}
    \rho \approx 1.24\,m_p\,[n_e(r)n_p(r)]^{1/2},
\end{eqnarray}
where $n_e$ and $n_p$ denote the number densities of electrons and protons, respectively. 
The product $n_e(r)n_p(r)$ can be expressed as
\begin{eqnarray}
    n_e(r)n_p(r)&=&\frac{(r/r_c)^{-\alpha'}}{(1+r^2/r_c^2)^{3\beta-\alpha'/2}}\frac{n_0^2}{(1+r^\gamma/r_s^\gamma)^{\varepsilon/\gamma}}\nonumber\\
    &&+\frac{n_0'^2}{(1+r^2/r_c'^2)^{2\beta'}}.\label{npne}
\end{eqnarray}
Table~\ref{parameterdensity} represents \textcolor{black}{all parameters needed in Eq.~(\ref{npne})}.
\begin{table*}[htbp]
\centering
\small
\caption{Parameters for the gas density profiles of the galaxy clusters. The data are taken from Ref.~\cite{Mashhoon2014}.}
\label{tab:cluster_params}
\begin{tabular}{lcccccccccc}
\hline\hline
Cluster 
& $r_{500}$ 
& $n_0$ 
& $r_c$ 
& $r_s$ 
& $\alpha'$ 
& $\beta$ 
& $\varepsilon$ 
& $n_0'$ 
& $r_c'$ 
& $\beta'$ \\
& (kpc) 
& ($10^{-3}\,\mathrm{cm^{-3}}$) 
& (kpc) 
& (kpc) 
&  
&  
&  
& ($10^{-1}\,\mathrm{cm^{-3}}$) 
& (kpc) 
&  \\
\hline
A133   & $1007 \pm 41$ & 4.705  & 94.6  & 1239.9 & 0.916 & 0.526 & 4.943 & 0.247 & 75.83 & 3.607 \\
A383   & $944 \pm 32$  & 7.226  & 112.1 & 408.7  & 2.013 & 0.577 & 0.767 & 0.002 & 11.54 & 1.000 \\
A478   & $1337 \pm 58$ & 10.170 & 155.5 & 2928.9 & 1.254 & 0.704 & 5.000 & 0.762 & 23.84 & 1.000 \\
A907   & $1096 \pm 30$ & 6.257  & 136.9 & 1887.1 & 1.556 & 0.594 & 4.998 & --    & --    & --    \\
A1413  & $1299 \pm 43$ & 5.239  & 195.0 & 2153.7 & 1.247 & 0.661 & 5.000 & --    & --    & --    \\
A1795  & $1235 \pm 36$ & 31.175 & 38.2  & 682.5  & 0.195 & 0.491 & 2.606 & 5.695 & 3.00  & 1.000 \\
A1991  & $732 \pm 33$  & 6.405  & 59.9  & 1064.7 & 1.735 & 0.515 & 5.000 & 0.007 & 5.00  & 0.517 \\
A2029  & $1362 \pm 43$ & 15.721 & 84.2  & 908.9  & 1.164 & 0.545 & 1.669 & 3.510 & 5.00  & 1.000 \\
A2390  & $1416 \pm 48$ & 3.605  & 308.2 & 1200.0 & 1.891 & 0.658 & 0.563 & --    & --    & --    \\
MKW4   & $634 \pm 28$  & 0.196  & 578.5 & 595.1  & 1.895 & 1.119 & 1.602 & 0.108 & 30.11 & 1.971 \\
\hline\hline
\end{tabular}
\label{parameterdensity}
\end{table*}

\textcolor{black}{In this sequence of calculations, each cluster is considered to be a spherical configuration of matter with an effective radius of $r_{500}$. Here, $r_{500}$ is defined to be the cluster radius within which the average overdensity is 500 times the critical density of the universe at the redshift of the cluster in the DM model~\cite{Mashhoon2014}.}

According to Ref.~\cite{Mashhoon2014}, the intracluster plasma can be divided into two regions: a cooling zone near the cluster center and an outer region. The temperature profiles in these two regions are modeled by two different functions. At the center of a cluster the temperature decreases due perhaps to radiative cooling in this region; hence
\begin{eqnarray}
    \tau_{in}(r)=\frac{x_0+T_{min}/T_0}{x_0+1},
\end{eqnarray}
where
\begin{eqnarray}
    x_0=\left(\frac{r}{r_{cool}}\right)^{a_{cool}}.
\end{eqnarray}
For the outside region of the central cooling zone,
\begin{eqnarray}
    \tau_{out}(r)=\frac{(r/r_t)^{a'}}{\left[1+\left(\frac{r}{r_t}\right)^b\right]^{c'/b}}.
\end{eqnarray}
Here, $r_t$ denotes the radial transision region. Temperature profile of the cluster reads
\begin{eqnarray}
    \tau(r)=T_0\tau_{in}(r)\tau_{out}(r).\label{tauuuu}
\end{eqnarray}
The parameters related to the temperature profile are given in Table~\ref{parametertemperature}
\begin{table*}[htbp]
\centering
\small
\caption{The parameters of the temperature profiles for ten galaxy clusters. The data are taken from Ref.~\cite{Mashhoon2014}.}
\begin{tabular}{lcccccccc}
\hline\hline
Cluster &
$T_0$ &
$r_t$ &
$\alpha'$ &
$b$ &
$c'$ &
$T_{\min}/T_0$ &
$r_{\mathrm{cool}}$ &
$\alpha_{\mathrm{cool}}$ \\
& (keV) & (Mpc) & & & & & (kpc) & \\
\hline
A133  & 3.61  & 1.42 &  0.12 & 5.00 & 10.0 & 0.27 &  57  & 3.88 \\
A383  & 8.78  & 3.03 & -0.14 & 1.44 &  8.0 & 0.75 &  81  & 6.17 \\
A478  & 11.06 & 0.27 &  0.02 & 5.00 &  0.4 & 0.38 & 129  & 1.60 \\
A907  & 10.19 & 0.24 &  0.16 & 5.00 &  0.4 & 0.32 & 208  & 1.48 \\
A1413 & 7.58  & 1.84 &  0.08 & 4.68 & 10.0 & 0.23 &  30  & 0.75 \\
A1795 & 9.68  & 0.55 &  0.00 & 1.63 &  0.9 & 0.10 &  77  & 1.03 \\
A1991 & 2.83  & 0.86 &  0.04 & 2.87 &  4.7 & 0.48 &  42  & 2.12 \\
A2029 & 16.19 & 3.04 & -0.03 & 1.57 &  5.9 & 0.10 &  93  & 0.48 \\
A2390 & 19.34 & 2.46 & -0.10 & 5.00 & 10.0 & 0.12 & 214  & 0.08 \\
MKW4  & 2.26  & 0.10 & -0.07 & 5.00 &  0.5 & 0.85 &  16  & 9.62 \\
\hline\hline
\end{tabular}
\label{parametertemperature}
\end{table*}

\textcolor{black}{In order to calculate the hydrostatic mass by using Eq.~(\ref{masstandardhyd}), we need the radial derivatives of the three-dimensional gas density and temperature profiles.} The entity $\frac{d\ln\rho}{d\ln r}$ in Eq.~(\ref{masstandardhyd}) is given by
\begin{strip}
\noindent\rule{\columnwidth}{0.4pt}  
\vspace{0.5em}
\begin{equation}
\begin{aligned}
\frac{d\ln \rho(r)}{d\ln r}
={}&
\frac{1}{H(r)}
\Bigg[
- n_0^{2}\alpha'
\left(\frac{r}{r_c}\right)^{-\alpha'}\left(1+\frac{r^{2}}{r_c^{2}}\right)^{0.5\alpha'-3\beta'}
\left(1+r^{\gamma}r_s^{-\gamma}\right)^{-\frac{\varepsilon}{\gamma}}+ n_0^{2}(\alpha'-6\beta)
\left(\frac{r}{r_c}\right)^{-\alpha'+2}
\left(1+\frac{r^{2}}{r_c^{2}}\right)^{-1+0.5\alpha'-3\beta}
\left(1+r^{\gamma}r_s^{-\gamma}\right)^{-\frac{\varepsilon}{\gamma}}
\\[1ex]
&- 6 n_0^{\prime 2}\beta'
\left(\frac{r}{r_c'}\right)^2
\left(1+\frac{r^{2}}{r_c^{\prime 2}}\right)^{-1-3\beta'
}- n_0^{2}\varepsilon
\left(\frac{r}{r_c}\right)^{-\alpha'}
\left(\frac{r}{r_s}\right)^{\gamma}
\left(1+\frac{r^{2}}{r_c^{2}}\right)^{0.5\alpha'-3\beta}
\left(1+r^{\gamma}r_s^{-\gamma}\right)^{-1-\frac{\varepsilon}{\gamma}}
\Bigg].\label{dlnrho}
\end{aligned}
\end{equation}
where
\begin{equation}
\begin{aligned}
H(r)
={}&
2\Bigg[
n_0^{\prime 2}
\left(1+\frac{r^{2}}{r_c^{\prime 2}}\right)^{-3\beta'}+
n_0^{2}
\left(\frac{r}{r_c}\right)^{-\alpha'}
\left(1+\frac{r^{2}}{r_c^{2}}\right)^{0.5\alpha'-3\beta}
\left(1+r^{\gamma}r_s^{-\gamma}\right)^{-\frac{\varepsilon}{\gamma}}
\Bigg].
\end{aligned}
\end{equation}
\textcolor{black}{Moreover, the entity $\frac{d\ln\tau}{d\ln r}$ can be written as}
\begin{equation}
    \begin{aligned}
        \textcolor{black}{\frac{d\ln \tau}{d\ln r}=a_{cool}\left(\frac{r}{r_{cool}}\right)^{a_{cool}}\frac{T_0-T_{min}}{[1+(\frac{r}{r_{cool}})^{a_{cool}}][T_{min}+T_0 (\frac{r}{r_{cool}})^{a_{cool}}]}-a'-\frac{c'\left(\frac{r}{r_t}\right)^b}{1+\left(\frac{r}{r_t}\right)^b}}.\label{dlntau}
    \end{aligned}
\end{equation}
\vspace{0.5em}
\noindent\rule{\textwidth}{0.4pt}  
\end{strip}

\textcolor{black}{More details on the standard calculation of hydrostatic mass can be referred to Refs.~\cite{Mashhoon2014,Vikhlinin2006}. The values of the parameters needed in the calculation of $\frac{d\ln\rho}{d\ln r}$ are presented in Table~\ref{parameterdensity}, while the ones needed in the calculation of $\frac{d\ln\tau}{d\ln r}$ are presented in Table~\ref{parametertemperature}.} 

\subsection{Hydrostatic mass of galaxy clusters in Rastall gravity}
\label{massrastall}
Rastall gravity is intrinsically formulated within a general relativistic framework, whereas the hydrostatic mass of galaxy clusters is commonly derived in the Newtonian limit. To bridge these two regimes, we first formulate the hydrostatic equilibrium condition in the general relativistic context, i.e. the TOV equation modified by Rastall gravity, and subsequently derive its weak-field limit. Throughout the general relativistic regime formulation, we adopt natural units by setting $G=c=1$. The physical constants will be reinstated when taking the Newtonian limit.

Consider an interior Schwarzschild metric of the form
\begin{equation}
ds^2 = -e^{2\nu(r)}dt^2 + e^{2\lambda(r)}dr^2 + r^2 d\theta^2 + r^2 \sin^2\theta d\phi^2,
\end{equation}
where $e^{2\lambda(r)} = \left(1 - \dfrac{2m(r)}{r}\right)^{-1}$. We also consider the EMT of a perfect fluid,
\begin{eqnarray}
T^\mu_{\ \nu} =
\begin{pmatrix}
-\rho {\quad} & 0 {\quad} & 0 {\quad} & 0 \\[0.8ex]
0 {\quad} & p {\quad} & 0 {\quad} & 0 \\[0.8ex]
0 {\quad} & 0 {\quad} & p {\quad} & 0 \\[0.8ex]
0 {\quad} & 0 {\quad} & 0 {\quad} & p
\end{pmatrix}.
\end{eqnarray}
By applying the Rastall field equations to the metric and the energy–momentum tensor given above, one obtains the Rastall-modified TOV equation, i.e.~\cite{Pattersons2025a}
\begin{eqnarray}
    \frac{dp}{dr}=-\frac{\frac{(\rho+p)\left(m+\frac{\kappa}{2}r^3\mathcal{P}\right)}{r(r-2m)}}{1-\frac{\kappa\lambda}{4\kappa\lambda-1}\left(3-\frac{d\rho}{dp}\right)},\label{TOVRastall}
\end{eqnarray}
where $\mathcal{P}=p-\left[\frac{\kappa\lambda}{4\kappa\lambda-1}(3p-\rho)\right]$.

Recall that, in GR, the hydrostatic equilibrium equation in the Newtonian regime can be recovered from the TOV equation by assuming $\rho \gg p$ and $r \gg m$, as has been mentioned before. By adopting the same procedure and also considering $\frac{d\rho}{dp}=\frac{\rho}{r}\frac{d\ln\rho}{d\ln r}\left(\frac{dp}{dr}\right)^{-1}$, we obtain the Newtonian-limit expression of Eq.~(\ref{TOVRastall}), i.e.
\begin{eqnarray}
    \frac{dp}{dr}=-\frac{\rho}{r^2}\left(\frac{m+\frac{1}{2}\frac{\kappa^2\lambda}{4\kappa\lambda-1}r^3\rho}{1-\frac{\kappa\lambda}{4\kappa\lambda-1}\left[3-\frac{\rho}{r}\frac{d\ln\rho}{d\ln r}\left(\frac{dp}{dr}\right)^{-1}\right]}\right).
\end{eqnarray}
With a little algebra, we find
\begin{eqnarray}
    \frac{dp}{dr}&=&\left[-\frac{\rho}{r^2}\left\{m+\frac{\kappa\lambda}{4\kappa\lambda-1}\left(\frac{\kappa r^3\rho}{2}+r\frac{d\ln\rho}{d\ln r}\right)\right\}\right]\nonumber\\
    &&\times\frac{4\kappa\lambda-1}{\kappa\lambda-1}.
\end{eqnarray}
Recall that for an ideal gas, we have Eq.~(\ref{idealgas}). Substituting this relation into the Rastall-modified Newtonian-limit hydrostatic equilibrium equation and restoring the gravitational constant $G$ and the speed of light $c$ in the expression, we obtain
\begin{eqnarray}
    m(r)&=&-\frac{k_B}{\mu m_p}\frac{\kappa\lambda-1}{4\kappa\lambda-1}\tau(r)r\left(\frac{d\ln\rho}{d\ln r}+\frac{d\ln\tau}{d\ln r}\right)\nonumber\\
    &&-\frac{\kappa\lambda}{4\kappa\lambda-1}\left(\frac{\kappa r^3\rho}{2}+\frac{c^2r}{G}\frac{d\ln\rho}{d\ln r}\right).\label{rastallhydrostaticmass}
\end{eqnarray}
Eq.~(\ref{rastallhydrostaticmass}) is the Rastall version of the hydrostatic mass of galaxy clusters. \textcolor{black}{The entities $\tau$, $\frac{d\ln\rho}{d\ln r}$, and $\frac{d\ln\tau}{d\ln r}$ can be calculated by using Eqs.~(\ref{tauuuu}), (\ref{dlnrho}), and (\ref{dlntau}), respectively.} When $\lambda=0$, the equation is reduced to Eq.~(\ref{recover}).

\begin{figure*}[!htbp]
\centering

\begin{minipage}{0.45\textwidth}
\centering
\includegraphics[width=\linewidth]{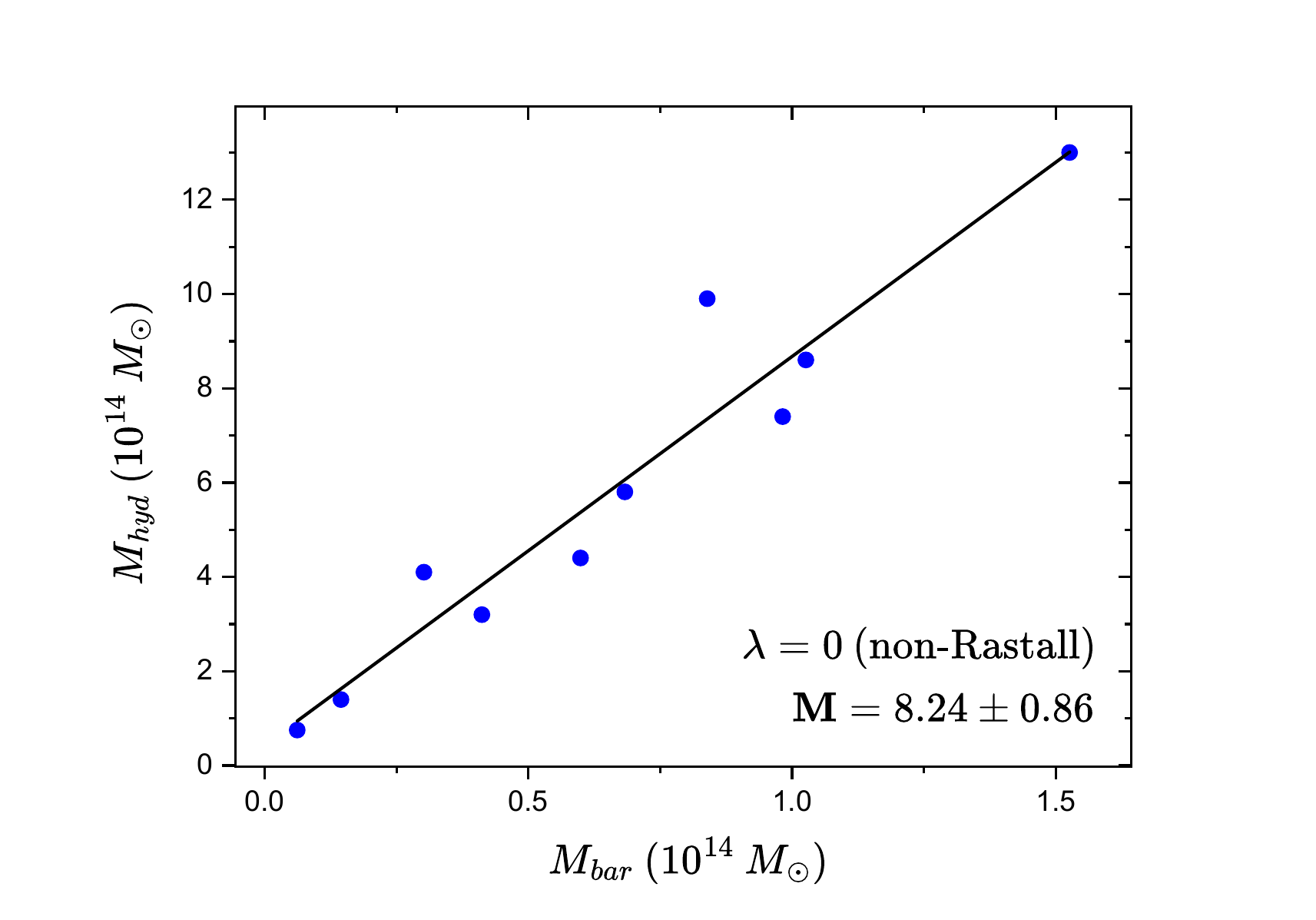}
\centering \textcolor{black}{(a)}
\end{minipage}
\hfill
\begin{minipage}{0.4\textwidth}
\centering
\includegraphics[width=\linewidth]{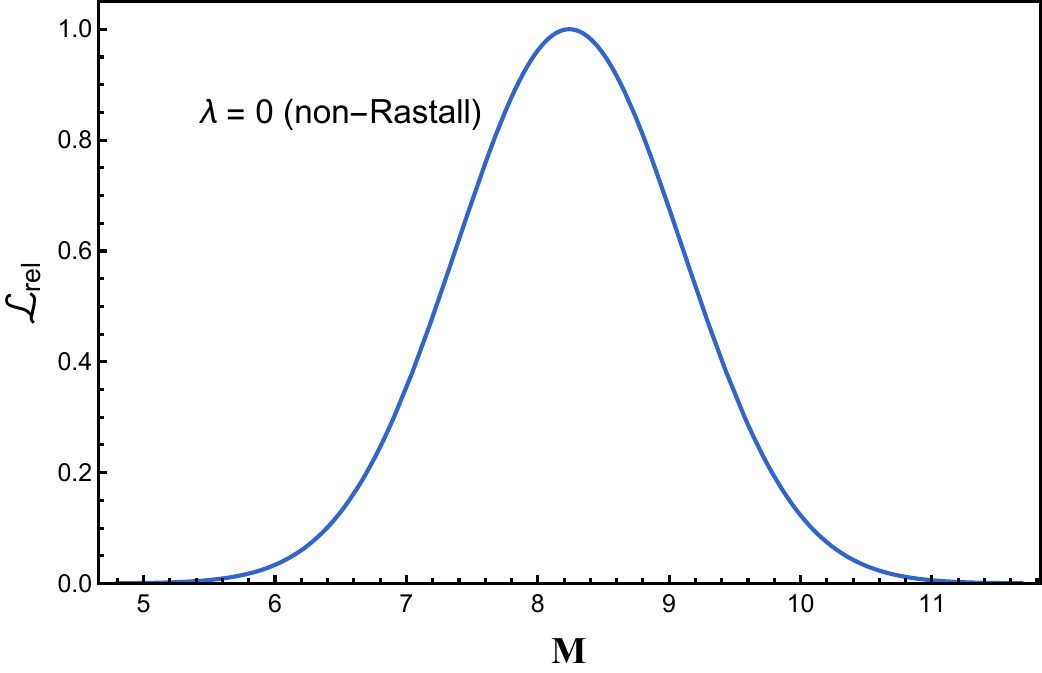}
\centering \textcolor{black}{(b)}
\end{minipage}

\vspace{0.3cm}

\begin{minipage}{0.45\textwidth}
\centering
\includegraphics[width=\linewidth]{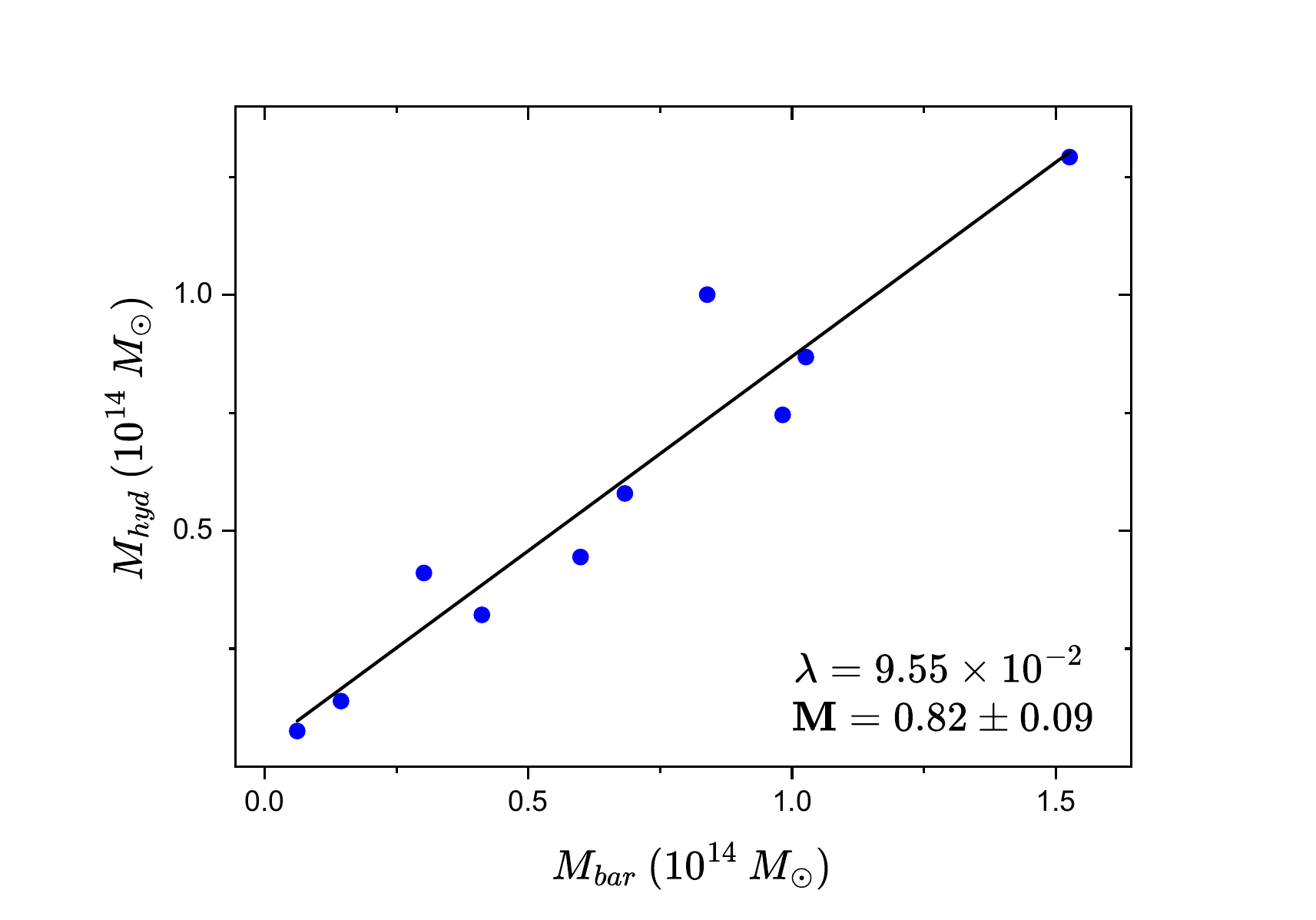}
\centering \textcolor{black}{(c)}
\end{minipage}
\hfill
\begin{minipage}{0.4\textwidth}
\centering
\includegraphics[width=\linewidth]{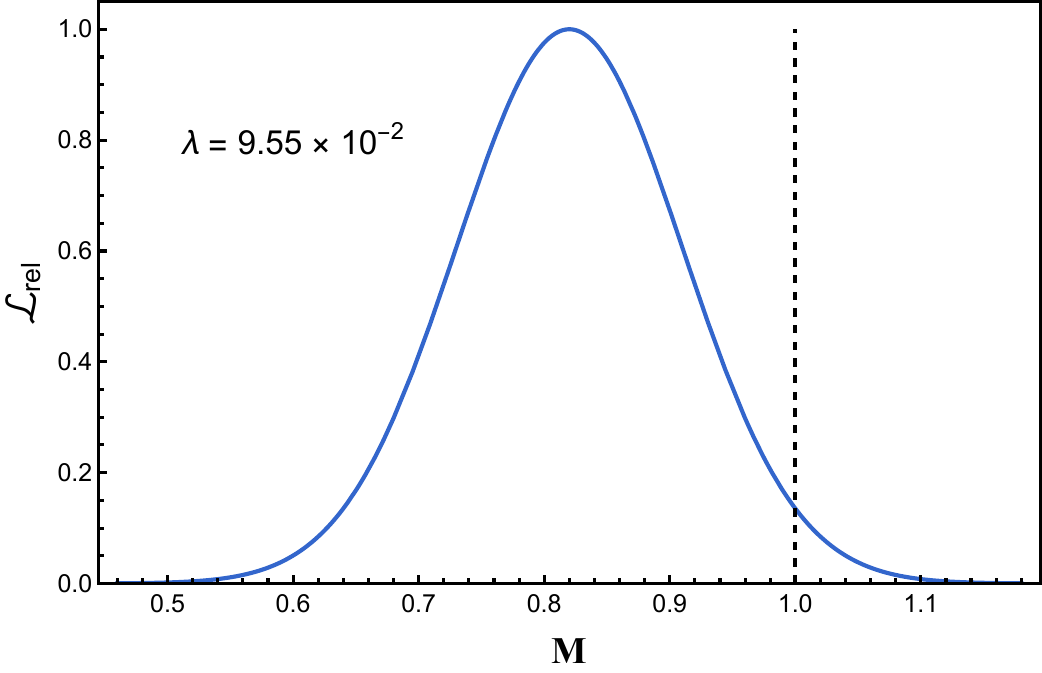}
\centering \textcolor{black}{(d)}
\end{minipage}

\vspace{0.3cm}

\begin{minipage}{0.45\textwidth}
\centering
\includegraphics[width=\linewidth]{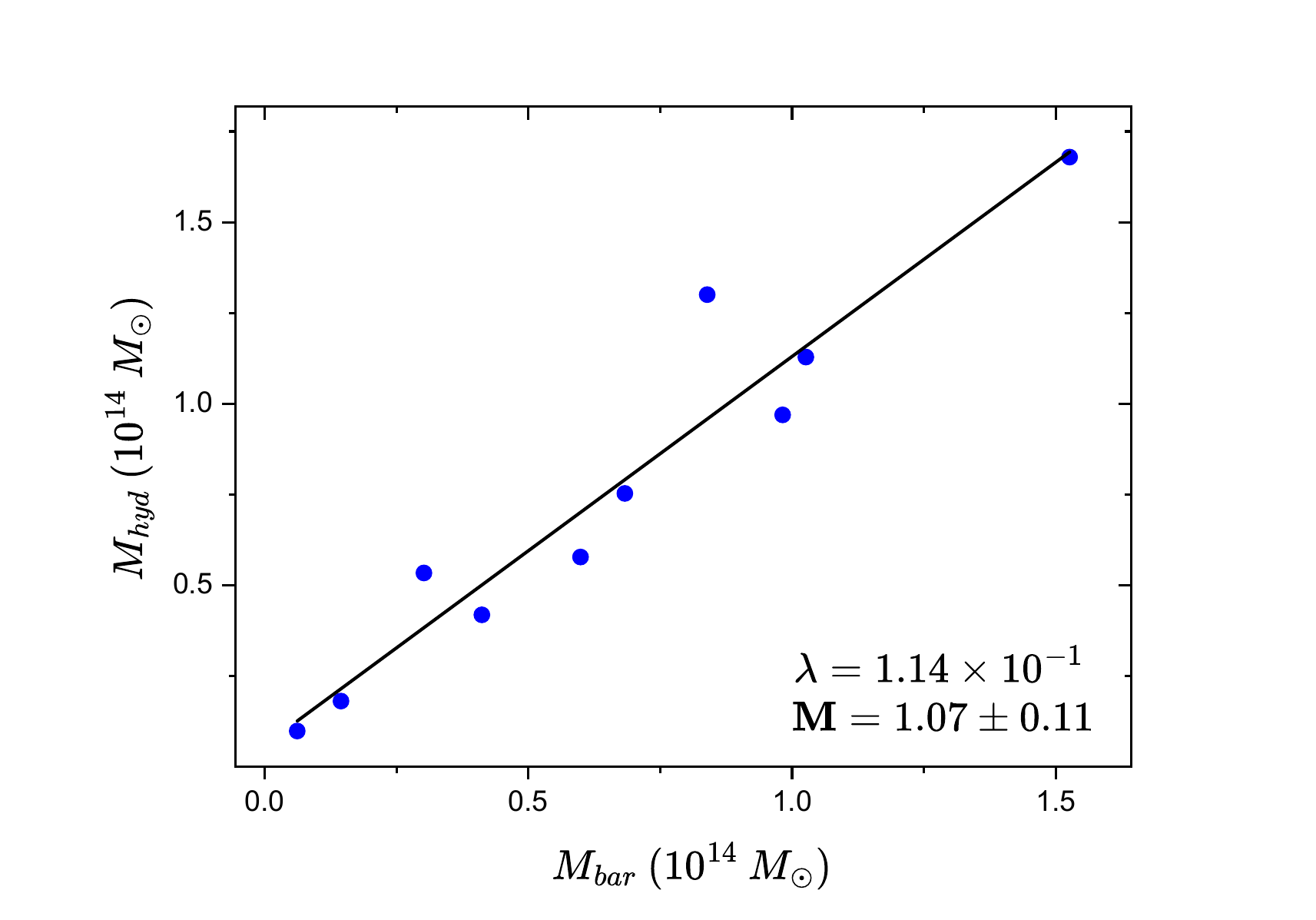}
\centering \textcolor{black}{(e)}
\end{minipage}
\hfill
\begin{minipage}{0.4\textwidth}
\centering
\includegraphics[width=\linewidth]{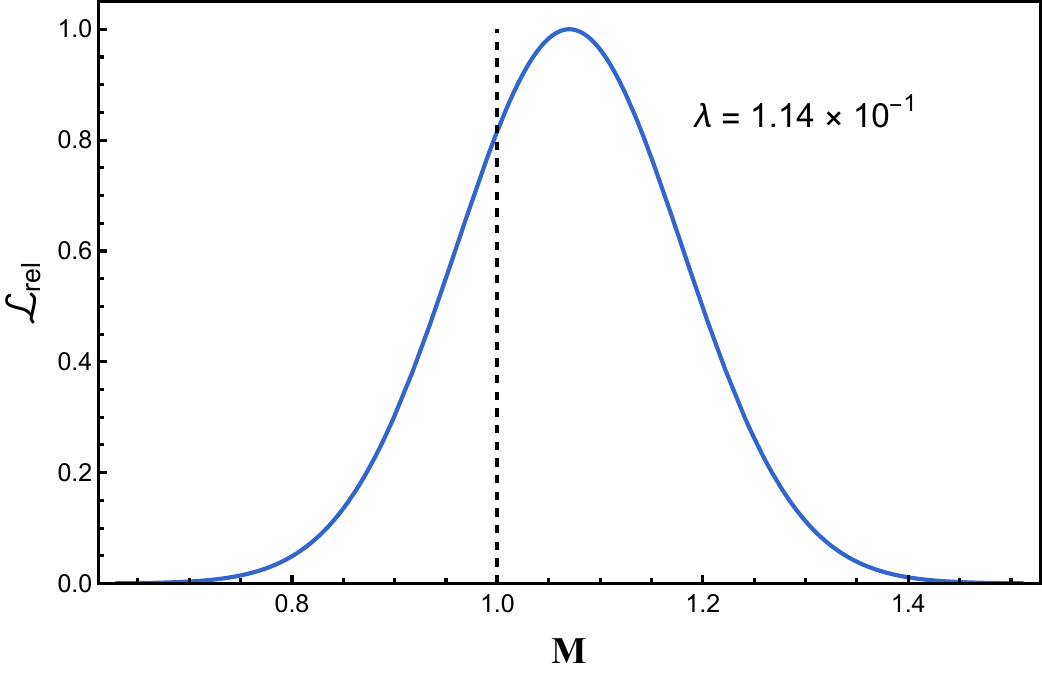}
\centering \textcolor{black}{(f)}
\end{minipage}

\caption{
(a) Linear fit to the relation between the hydrostatic masses according to standard non-Rastall gravity and the observed baryonic masses, (b) the likelihood function for parameter $\mathbf{M}$ in standard non-Rastall gravity, (c) linear fit to the relation between the hydrostatic masses according to Rastall gravity with $\lambda=9.55\times10^{-2}$ and the observed baryonic masses, (d) the likelihood function for parameter $\mathbf{M}$ in Rastall gravity for $\lambda=9.55\times10^{-2}$, (e) linear fit to the relation between the hydrostatic masses according to Rastall gravity with $\lambda=1.14\times10^{-1}$ and the observed baryonic masses, and (f) the likelihood function for parameter $\mathbf{M}$ in Rastall gravity for $\lambda=1.14\times10^{-1}$.
}
\label{mbar}
\end{figure*}

\begin{figure*}[!htbp]
\centering

\begin{minipage}{0.45\textwidth}
\centering
\includegraphics[width=\linewidth]{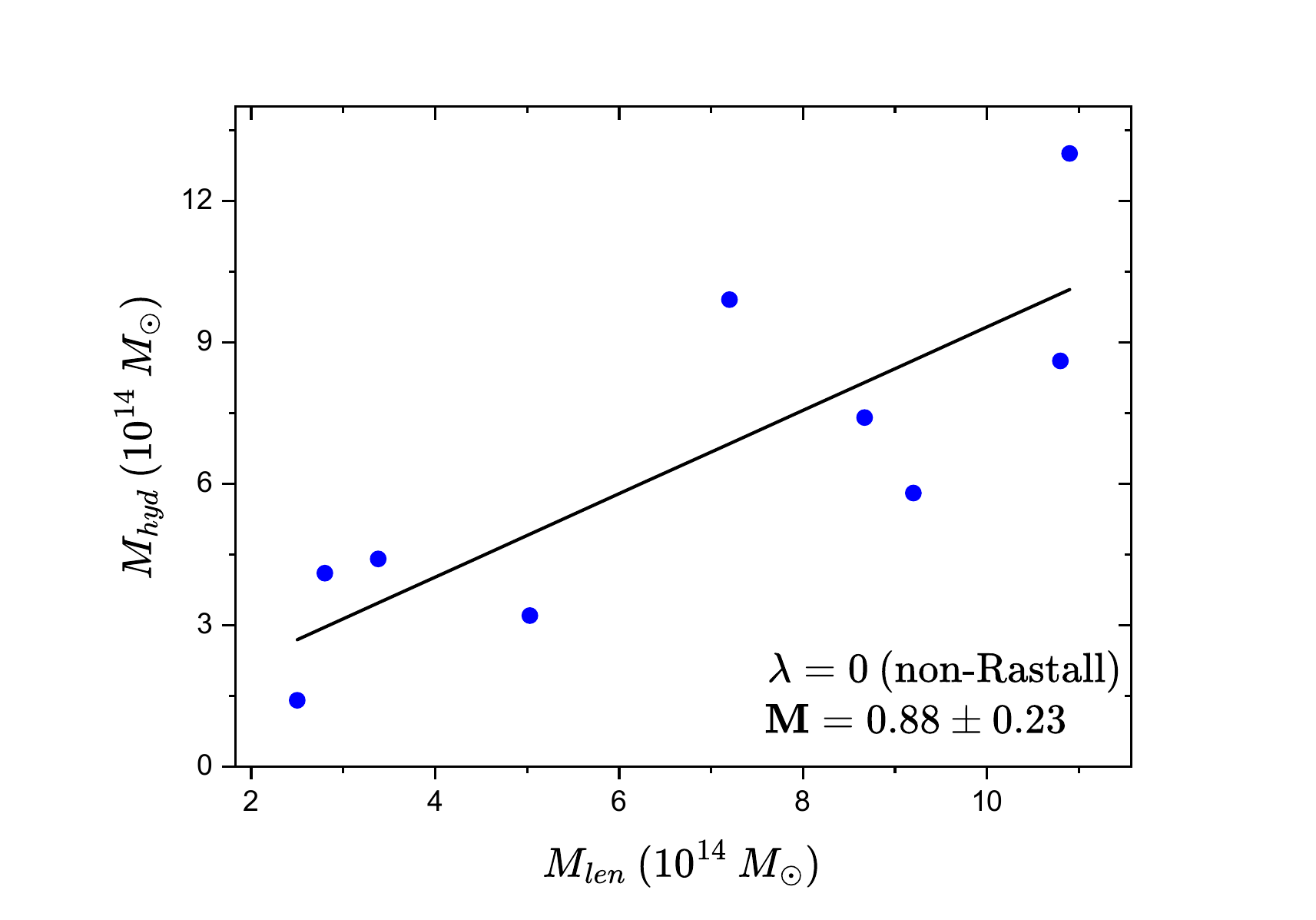}
\centering \textcolor{black}{(a)}
\end{minipage}
\hfill
\begin{minipage}{0.4\textwidth}
\centering
\includegraphics[width=\linewidth]{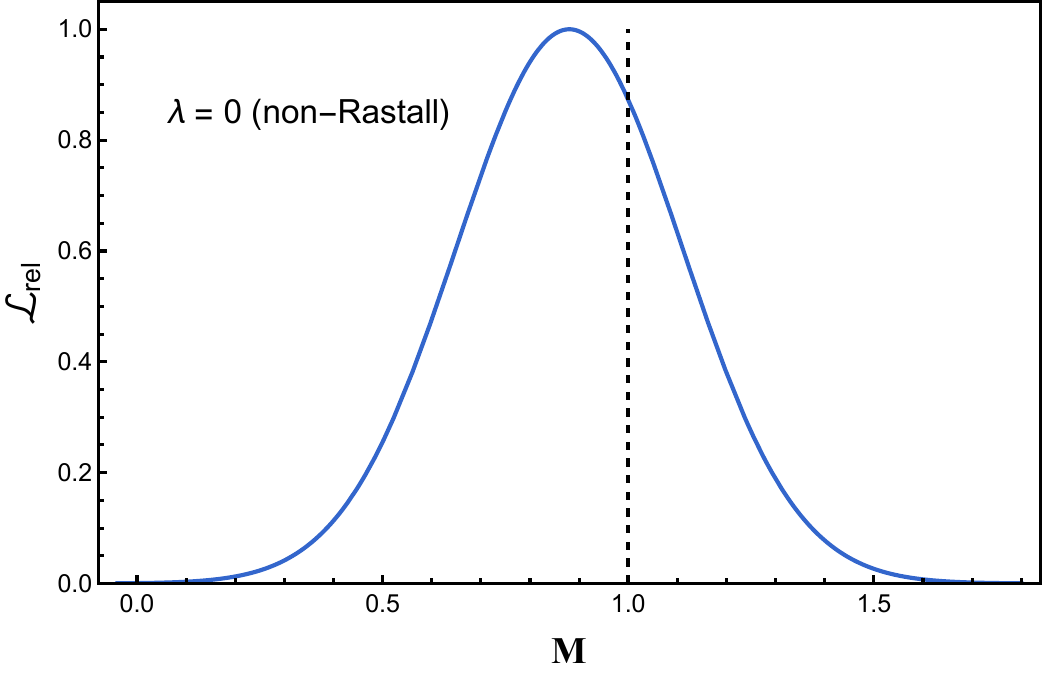}
\centering \textcolor{black}{(b)}
\end{minipage}

\vspace{0.3cm}

\begin{minipage}{0.45\textwidth}
\centering
\includegraphics[width=\linewidth]{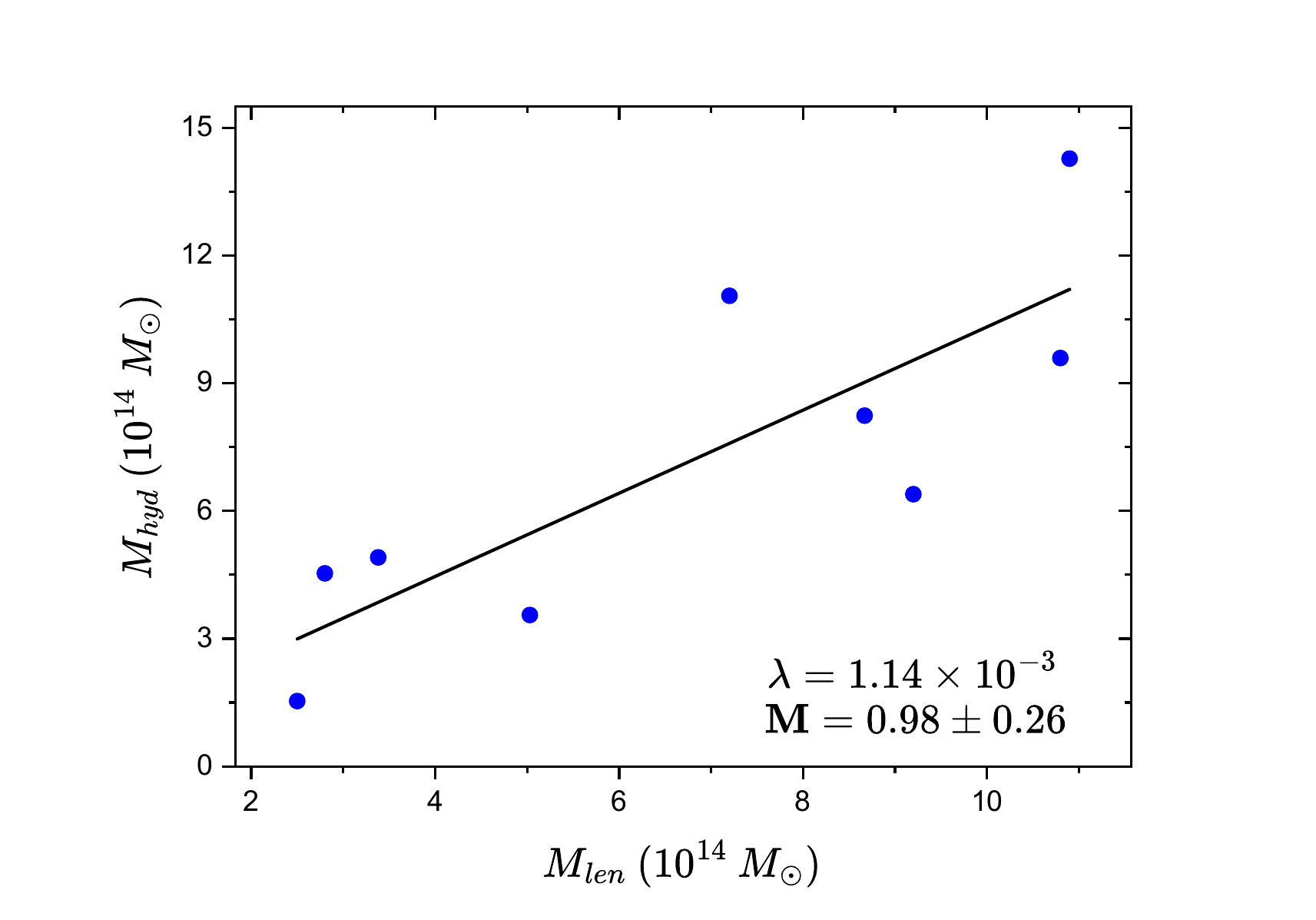}
\centering \textcolor{black}{(c)}
\end{minipage}
\hfill
\begin{minipage}{0.4\textwidth}
\centering
\includegraphics[width=\linewidth]{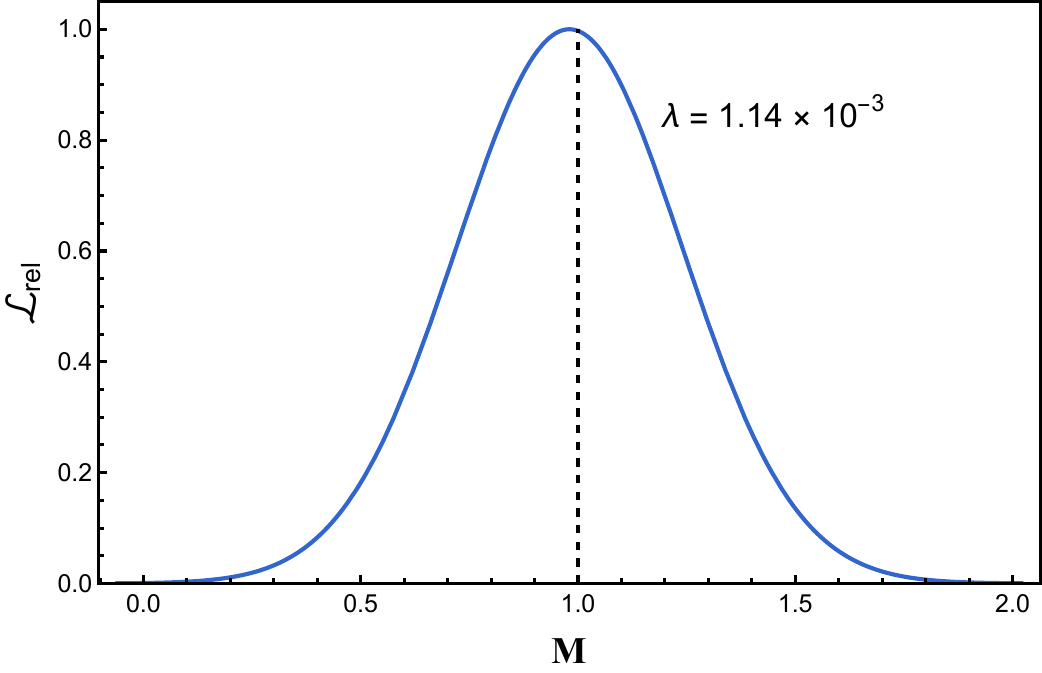}
\centering \textcolor{black}{(d)}
\end{minipage}

\vspace{0.3cm}

\begin{minipage}{0.45\textwidth}
\centering
\includegraphics[width=\linewidth]{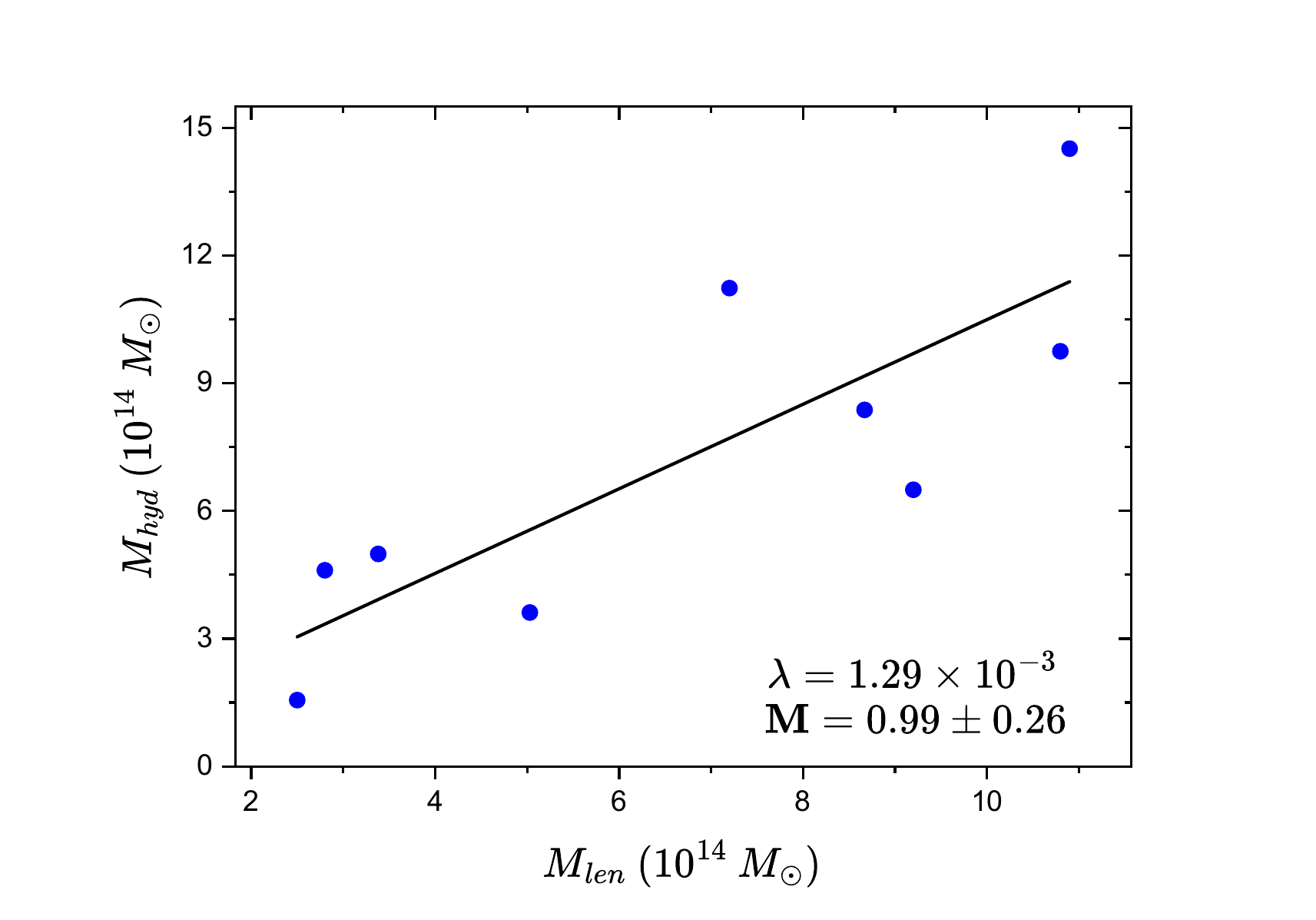}
\centering \textcolor{black}{(e)}
\end{minipage}
\hfill
\begin{minipage}{0.4\textwidth}
\centering
\includegraphics[width=\linewidth]{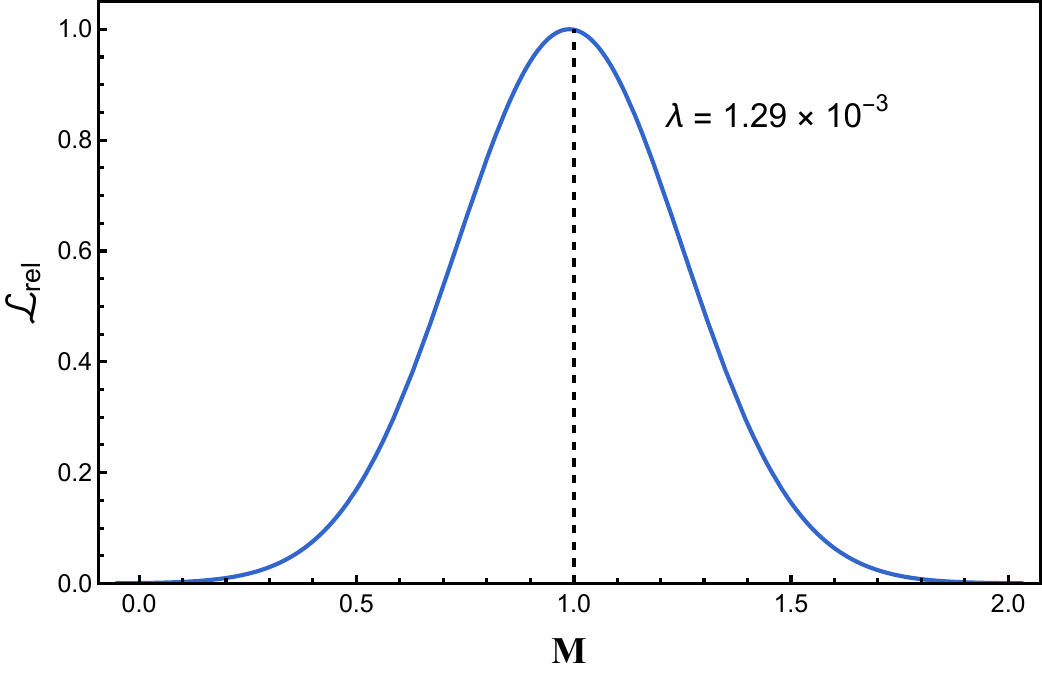}
\centering \textcolor{black}{(f)}
\end{minipage}

\caption{
\textcolor{black}{(a) Linear fit to the relation between the hydrostatic masses according to standard non-Rastall gravity and the observed lensing masses, (b) the likelihood function for parameter $\mathbf{M}$ in standard non-Rastall gravity, (c) linear fit to the relation between the hydrostatic masses according to Rastall gravity with $\lambda=1.14\times10^{-3}$ and the observed baryonic masses, (d) the likelihood function for parameter $\mathbf{M}$ in Rastall gravity for $\lambda=1.14\times10^{-3}$, (e) linear fit to the relation between the hydrostatic masses according to Rastall gravity with $\lambda=1.29\times10^{-3}$ and the observed baryonic masses, and (f) the likelihood function for parameter $\mathbf{M}$ in Rastall gravity for $\lambda=1.29\times10^{-3}$.
}}
\label{mlen}
\end{figure*}

\section{\textcolor{black}{Numerical results, statistical evaluations, and discussion}}
\label{numerical}

\begin{table*}
\centering
\small
\caption{Comparison between the Newtonian hydrostatic mass, the hydrostatic mass in Rastall gravity for two different values of the Rastall parameter $\lambda$, and the observed baryonic mass of galaxy clusters. The Newtonian hydrostatic mass and $r_{500}$ data are taken from Ref.~\cite{Apryandi2025}, while the baryonic mass data are taken from Ref.~\cite{Mashhoon2014}.}
\begin{tabular}{lccccc}
\hline\hline
Cluster & $r_{500}$ & $M_{\mathrm{Newt}}$ & $M_{\mathrm{hyd}}^{\mathrm{R}}$ & $M_{\mathrm{hyd}}^{\mathrm{R}}$ & $M_{\mathrm{bar}}$ \\
 & (kpc) & $(10^{14} M_\odot)$ & $(10^{14} M_\odot)$ & $(10^{14} M_\odot)$ & $(10^{14} M_\odot)$ \\
 & & & $\lambda = 9.55\times10^{-2}$ & $\lambda = 1.14\times10^{-1}$ & \\
\hline
A133 & $1007 \pm 41$ & 4.10 & 0.41 & 0.53 & $0.302 \pm 0.054$ \\
A383 & $944 \pm 32$ & 3.20 & 0.32 & 0.42 & $0.412 \pm 0.059$ \\
A478 & $1337 \pm 58$ & 7.40 & 0.75 & 0.97 & $0.982 \pm 0.204$ \\
A907 & $1096 \pm 30$ & 4.40 & 0.44 & 0.58 & $0.599 \pm 0.072$ \\
A1413 & $1299 \pm 43$ & 9.90 & 1.00 & 1.30 & $0.839 \pm 0.134$ \\
A1795 & $1235 \pm 36$ & 5.80 & 0.58 & 0.75 & $0.683 \pm 0.090$ \\
A1991 & $732 \pm 33$ & 1.40 & 0.14 & 0.18 & $0.145 \pm 0.027$ \\
A2029 & $1362 \pm 43$ & 8.60 & 0.87 & 1.13 & $1.026 \pm 0.146$ \\
A2390 & $1416 \pm 48$ & 13.00 & 1.29 & 1.68 & $1.526 \pm 0.246$ \\
MKW4 & $634 \pm 28$ & 0.75 & 0.08 & 0.10 & $0.062 \pm 0.008$ \\
\hline\hline
\end{tabular}
\label{tablembar}
\end{table*}

In this work, we compute the masses of galaxy clusters at the radius $r_{500}$. 
As shown in Tables~\ref{tablembar} and \ref{tablemlen}, the values of $r_{500}$ 
are subject to observational uncertainties. In our analysis, however, the mass 
is evaluated at the central (best-fit) value of $r_{500}$, while the propagation 
of the uncertainty in $r_{500}$ is not taken into account. This approach is also employed in Ref.~\cite{Apryandi2025}.

\begin{table*}
\centering
\small
\caption{Comparison between the hydrostatic Newtonian mass, hydrostatic mass in Rastall gravity for two values of the Rastall parameter, and the lensing mass of galaxy clusters. The Newtonian mass and $r_{500}$ data are taken from Ref.~\cite{Apryandi2025}, while the lensing mass data are taken from Refs.~\cite{Mahdavi2008,Okabe2016,Herbonnet2020,Nagarajan2019}.}
\begin{tabular}{lccccc}
\hline\hline
Cluster & $r_{500}$ & $M_{\mathrm{Newt}}$ & $M_{hyd}^{\text{R}}$ & $M_{hyd}^{\text{R}}$ & $M_{\mathrm{len}}$ \\
 & (kpc) & $(10^{14} M_\odot)$ & $(10^{14} M_\odot)$ & $(10^{14} M_\odot)$ & $(10^{14} M_\odot)$ \\
 & & & \textcolor{black}{$\lambda = 1.14\times10^{-3}$} & \textcolor{black}{$\lambda = 1.29\times10^{-3}$} & \\
\hline
\textcolor{black}{A133} & \textcolor{black}{$1007\pm41$} & \textcolor{black}{4.10} & \textcolor{black}{4.61} & \textcolor{black}{4.53} & \textcolor{black}{$2.8\pm 1.4$}\\
A383 & $944 \pm 32$ & 3.20 & \textcolor{black}{3.55} & \textcolor{black}{3.61} & 5.03 \textcolor{black}{$\pm1.34$} \\
A478 & $1337 \pm 58$ & 7.40 & \textcolor{black}{8.24} & \textcolor{black}{8.37} & 8.67 \textcolor{black}{$\pm2.01$} \\
\textcolor{black}{A907} & \textcolor{black}{$1096\pm30$} & \textcolor{black}{4.40} & \textcolor{black}{4.91} & \textcolor{black}{4.99} & \textcolor{black}{$3.38\pm0.94$}\\
\textcolor{black}{A1413} & \textcolor{black}{$1299\pm43$} & \textcolor{black}{$9.90$} & \textcolor{black}{11.05} & \textcolor{black}{11.23} & \textcolor{black}{$7.20\pm2.00$} \\
A1795 & $1235 \pm 36$ & 5.80 & \textcolor{black}{6.39} & \textcolor{black}{6.50} & 9.20 \textcolor{black}{$\pm2.3$} \\
A1991 & $732 \pm 33$ & 1.40 & \textcolor{black}{1.53} & \textcolor{black}{1.56} & 2.50 \textcolor{black}{$\pm1.75$} \\
A2029 & $1362 \pm 43$ & 8.60 & \textcolor{black}{9.59} & \textcolor{black}{9.75} & 10.80 \textcolor{black}{$\pm2.10$} \\
\textcolor{black}{A2390} & \textcolor{black}{$1416\pm48$} & \textcolor{black}{13.00} & \textcolor{black}{14.27} & \textcolor{black}{14.51} & \textcolor{black}{$10.9\pm1.7$}\\
\hline\hline
\end{tabular}
\label{tablemlen}
\end{table*}

We consider two distinct possibilities. The first possibility assumes the absence of DM; in this case, we fit the hydrostatic mass derived within the framework of Rastall gravity to the observed baryonic mass in order to obtain the best linear fit. The ideal situation corresponds to a slope $\mathbf{M}=1$, which indicates a one-to-one correspondence between the hydrostatic and baryonic masses. This approach has been adopted in Ref.~\cite{Apryandi2025} using EiBI theory, BHG, and Newtonian gravity modified by the GUP; in Ref.~\cite{Mashhoon2014} within the context of nonlocal gravity; and in Ref.~\cite{Moffat2014} using STVG.

The second possibility assumes the existence of DM, but aims to alleviate the mass discrepancy between the hydrostatic and the gravitational lensing masses. In this scenario, we perform a linear fit between the hydrostatic mass in Rastall gravity and the observed lensing mass to obtain the best-fit slope. Similarly to the first case, the ideal situation is achieved when the slope $\mathbf{M}=1$. This approach has been employed in Ref.~\cite{Desai2022}, where an effective mass derived from the Kottler metric was used.

Table~\ref{tablembar} presents a comparison between the Newtonian hydrostatic mass, the hydrostatic mass in Rastall gravity for two values of the Rastall parameter, namely $\lambda = 9.55\times10^{-2}$ and $\lambda = 1.14\times10^{-1}$, and the observed baryonic mass of galaxy clusters. Here we consider ten galaxy clusters as the sample. From the values listed in the table, it is evident that the Rastall parameter can reduce the hydrostatic mass in such a way that it approaches the observed baryonic mass.

\textcolor{black}{Since our aim in the first scenario is to constrain the Rastall parameter within the hydrostatic mass framework under the assumption of no DM, we consider parameter values that yield the best agreement with the observational data. In particular, we focus on two representative values of the Rastall parameter that produce the closest approach to a linear relation with slope near unity. While one of these values provides the overall best-fit, the second value is also retained for comparison purposes, in order to illustrate the sensitivity of the results to variations of the Rastall parameter}

\textcolor{black}{To further quantify the consistency between the model and the data, we also evaluate the corresponding relative likelihood.} \textcolor{black}{Note that since we are working with Rastall gravity that results in extra degree of freedom brought by Rastall parameter $\lambda$, and the uncertainty always appears in the measurements, we also have to evaluate the goodness-of-fit by calculating the chi-squared $\chi^2$ and reduced chi-squared $\chi_{\nu}^2$ of every fit.}

In this work, the relative likelihood\textcolor{black}{\footnote{\textcolor{black}{
The relative likelihood used in this work is not fully rigorous, as it only accounts for statistical uncertainty and does not incorporate systematic uncertainty, since the error bars of the observational data are not included in the analysis. Therefore, a more robust approach would involve constructing the likelihood based on the $\chi^2$ statistics and performing a full Markov Chain Monte Carlo~(MCMC) analysis.
}}} is calculated using the Gaussian form~\cite{Mashhoon2014}
\begin{eqnarray}
    \textcolor{black}{\mathcal{L}_{\mathrm{rel}} =
    \exp\left[
    -\frac{1}{2}
    \left(
    \frac{\mathbf{M}-\mu}{\sigma}
    \right)^2
    \right],}\label{likelihoodrumus}
\end{eqnarray}
\textcolor{black}{where $\mathbf{M}$ denotes the ratio $M_{hyd}/M_{bar}$, $\mu$ is the slope obtained from the linear fit, and $\sigma$ is the uncertainty associated with the fit.}

\textcolor{black}{The formula of $\chi^2$ is given by~\cite{Andrae}}
\begin{eqnarray}
    \textcolor{black}{\chi^2=\sum^N_{n=1}\left(\frac{y_n-E_n}{\sigma_n}\right),}\label{chirumus}
\end{eqnarray}
\textcolor{black}{where we have $N$ data values $y_n$, measured with uncertainties $\sigma_n$, and $E_n$ denotes the value from the model. Moreover, $\chi_\nu^2$ reads}
\begin{eqnarray}
    \textcolor{black}{\chi_\nu^2=\frac{\chi^2}{N-\mathscr{P}}.}\label{reducedchirumus}
\end{eqnarray}
\textcolor{black}{Here $\mathscr{P}$ denotes the amount of fitted parameters.}

From Fig.~\ref{mbar}\textcolor{black}{(a)}, it is evident that the standard (non-Rastall) hydrostatic mass yields a slope that deviates significantly from unity, namely $\mathbf{M}=8.24 \pm 0.86$. Even when the uncertainty is taken into account, the value $\mathbf{M}=1$ lies well outside the allowed range. \textcolor{black}{Therefore, the strong mismatch with the observed baryonic mass under the assumption of no DM reflects the well-known discrepancy in which the hydrostatic mass substantially exceeds the baryonic contribution. Moreover, from Fig.~\ref{mbar}(b), it is evident that the value $\mathbf{M}=1$ does not lie within the tail of the likelihood distribution, indicating that the standard Newtonian estimate is strongly disfavored by the observed baryonic mass data.}

\textcolor{black}{In this calculation, we found that the values $\lambda=9.55\times10^{-2}$ and $\lambda = 1.14\times10^{-1}$ result in the slopes which are very close to unity, as can be seen in Fig.~\ref{mbar}(c) and (e). In Fig.~\ref{mbar}(d) and (f), it is obvious that the value $\mathbf{M}=1$ lies within the curve of the likelihood distribution, which means that both values of parameter are favored by the observational baryonic data.} For Rastall gravity with $\lambda = 9.55\times10^{-2}$, the resulting slope is significantly improved compared to the non-Rastall case. \textcolor{black}{As we can see in Fig.~\ref{mbar}(d),} although the value $\mathbf{M}=1$ is still not encompassed by the uncertainty of the slope, it is nevertheless included within the corresponding relative likelihood curve.

In the comparison between the hydrostatic mass and the baryonic mass, the value $\lambda = 1.14\times10^{-1}$ yields the best linear fit, with a slope $\mathbf{M} = 1.07 \pm 0.11$, \textcolor{black}{as shown by Fig.~\ref{mbar}(e)}. The slope is very close to unity, and the value $\mathbf{M}=1$ lies within both the uncertainty range and the corresponding relative likelihood curve, \textcolor{black}{as shown by Fig.~\ref{mbar}(f)}. This result indicates that, from a statistical standpoint, Rastall gravity provides a more favorable description of the hydrostatic-baryonic mass relation than standard gravity under the assumption of the absence of DM. Another noteworthy result is that the best-fit slope obtained in this work is closer to unity than those reported in some related studies. In Ref.~\cite{Apryandi2025}, the best-fit slopes were found to be $0.126 \pm 0.086$ for EiBI theory and $0.023 \pm 0.119$ for BHG. Meanwhile, Ref.~\cite{Mashhoon2014} reported a slope of $0.84 \pm 0.04$ within the framework of nonlocal gravity.

In the second scenario, we also examine whether the hydrostatic mass within the framework of Rastall gravity can help to alleviate the hydrostatic mass bias, an issue that was not resolved in Ref.~\cite{Desai2022}. To this end, we consider a sample of five galaxy clusters. The values of the Newtonian hydrostatic mass, the Rastall hydrostatic mass, and the corresponding lensing mass for these clusters are presented in Table~\ref{tablemlen}. Note that in Table~\ref{tablemlen}, the measurement uncertainties are treated as symmetric. In the original dataset, some uncertainties are asymmetric. To symmetrize them, we take the mean values of the upper and lower uncertainties. \textcolor{black}{As in the first scenario, the aim at this stage is to constrain the Rastall parameter in the cluster-scale, now under the assumption of the presence of DM, while examining whether the discrepancy between hydrostatic and lensing mass estimates can be alleviated. We consider representative parameter values that yield good agreement with the observational data.} For this test, we adopt $\lambda = 1.49\times10^{-3}$ and $\lambda = 1.69\times10^{-3}$. Both values lead to an increase in the hydrostatic mass \textcolor{black}{and result in slopes close to unity}.

The linear fits between the standard (non-Rastall) hydrostatic mass and the observed lensing mass, as well as between the Rastall hydrostatic mass and the observed lensing mass, are shown in Fig.~\ref{mlen}. The standard non-Rastall hydrostatic mass yields a slope value of \textcolor{black}{$\mathbf{M}=0.88\pm0.23$}, \textcolor{black}{as shown by Fig.~\ref{mlen}(a)}. Although this value is already close to unity, we further investigate whether the inclusion of Rastall gravity can improve the linear relation and bring the slope closer to the ideal value $\mathbf{M}=1$, thereby alleviating the hydrostatic mass bias.

In the comparison between the hydrostatic mass and the lensing mass, the value \textcolor{black}{$\lambda = 1.14\times10^{-3}$} yields the  slope \textcolor{black}{$\mathbf{M} = 0.98 \pm 0.26$}, \textcolor{black}{as shown by Fig.~\ref{mlen}(c)}. Furthermore, the value \textcolor{black}{$\lambda = 1.29\times10^{-3}$} yields the best linear fit, with a slope \textcolor{black}{$\mathbf{M} = 0.99 \pm 0.26$}. The slope is very close to unity, even closer to $\mathbf{M}=1$ than the best-fit value obtained from the comparison between the hydrostatic mass and the observed baryonic mass. Moreover, the value $\mathbf{M}=1$ lies within the uncertainty range and is also located very close to the peak of the corresponding relative likelihood curve, \textcolor{black}{as we can see in Figs.~\ref{mlen}(b), (d), (f)}.

\begin{table}[h]
\label{tablechi}
\centering
{\color{black}
\caption{\textcolor{black}{Comparison of the goodness-of-fit for different values of the Rastall parameter $\lambda$.}}
\begin{tabular}{c c c c}
\hline
$\lambda$ & $\chi^2$ & $\chi_\nu^2$ & observed mass \\
\hline
$0$ & 33184.60 & 3318.46 & baryonic mass \\
$9.55\times10^{-2}$ & 22.51 & 2.50 & baryonic mass \\
$1.14\times10^{-1}$ & 55.45 & 6.16 & baryonic mass \\
$0$ & 11.33 & 1.26 & lensing mass \\
$1.14\times10^{-3}$ & 15.35 & 1.92 & lensing mass \\
$1.29\times10^{-3}$ & 16.09 & 2.01 & lensing mass \\
\hline
\label{tablechi}
\end{tabular}
}

\end{table}

\begin{table}[h]
\centering
{\color{black}
\caption{\textcolor{black}{Comparison of the goodness-of-fit for Rastall gravity and non-local gravity.}}
\begin{tabular}{c c c }
\hline
modified gravity & $\chi^2$ & $\chi_\nu^2$ \\
\hline
Rastall gravity, $\lambda=9.55\times10^{-2}$ & 22.51 & 2.50  \\
Rastall gravity, $\lambda=1.14\times10^{-1}$ & 55.45 & 6.16 \\
non-local gravity & 15.62 & 1.74 \\
\hline
\label{tablechicompare}
\end{tabular}
}
\end{table}

This result indicates that, \textcolor{black}{in the context of the linear fit}, Rastall gravity is able to significantly alleviate the hydrostatic mass bias. Nevertheless, a test with a larger sample of galaxy clusters and more lensing mass data is necessary, both to further assess the viability of Rastall gravity and to \textcolor{black}{obtain the more reliable constraint of} the value of the Rastall parameter $\lambda$ using observational data. Another noteworthy outcome of this study is that it addresses the issue encountered in Ref.~\cite{Desai2022}, where the hydrostatic mass bias could not be alleviated within the considered framework.

\textcolor{black}{Since there exist uncertainties in the measurements, as mentioned before, it is necessary to evaluate the goodness-of-fit using $\chi^2$ and $\chi^2_\nu$ based on Eqs.~(\ref{chirumus}) and (\ref{reducedchirumus}), respectively. It is worth noting that larger values of $\chi^2$ and $\chi^2_\nu$ indicate that the measurements deviate more strongly from the values predicted by the model. Moreover, the denominator in the $\chi^2_\nu$ evaluation contains the number of parameters, which is relevant for Rastall gravity as it introduces an additional degree of freedom, namely $\lambda$.}

\textcolor{black}{The results of the goodness-of-fit calculation are presented in Table~\ref{tablechi}. It can be seen that the improvement in the linear fit (i.e., the slope approaching unity) does not always correspond to a better goodness-of-fit. In particular, although certain values of the Rastall parameter yield slopes that are closer to the expected value, the corresponding $\chi_\nu^2$ values indicate that the overall agreement with the data is not necessarily improved.}

\textcolor{black}{This behavior suggests that, while Rastall gravity is able to capture the global trend of the data (as reflected in the slope), there remain non-negligible discrepancies at the level of individual data points. In other words, the model can improve the overall scaling relation, but does not fully account for the scatter in the observational data.}

\textcolor{black}{A similar feature is observed in the second scenario, where the comparison with lensing mass shows that the inclusion of Rastall gravity does not always lead to a better goodness-of-fit compared to the standard (non-Rastall) case. This indicates that, although Rastall gravity can alleviate the hydrostatic mass bias in terms of the slope, it does not systematically improve the statistical consistency with the data.}

\textcolor{black}{Note that, in the context of the first scenario where we confront the Rastall version of the hydrostatic mass of galaxy clusters with the observed baryonic mass, we use the dataset of the observed baryonic mass employed in Ref.~\cite{Mashhoon2014}. In that work, non-local gravity is used as the modified theory of gravity, and the authors obtain the best-fit value of $\mathbf{M} = 0.84 \pm 0.04$, which is lower than our best-fit value in the first scenario, i.e., $\mathbf{M} = 1.07 \pm 0.11$. However, in their work, the goodness-of-fit was not evaluated. This motivates us to calculate their goodness-of-fit and compare it with our results.}

\textcolor{black}{As shown in Table~\ref{tablechicompare}, the comparison with non-local gravity reveals that the latter provides a better goodness-of-fit, despite the fact that the linear relation obtained in the Rastall framework is closer to the expected one. This highlights a trade-off between capturing the global behavior of the data and achieving a statistically better fit.}

\textcolor{black}{Overall, these results indicate that Rastall gravity provides a viable phenomenological framework that can improve certain aspects of the mass discrepancy problem, particularly at the level of scaling relations. However, it does not universally outperform other modified gravity models, such as non-local gravity, when evaluated using standard goodness-of-fit criteria. This suggests that further investigations, including larger data samples and more refined modeling, are necessary to fully assess its effectiveness.}

\section{Conclusion}
\label{conclusion}
In this study, we have derived the formalism of the hydrostatic mass of galaxy clusters within the framework of Rastall gravity. We find that Rastall gravity has a nontrivial influence on the hydrostatic mass of galaxy clusters. Two scenarios are considered: (i) the absence of DM, in which the Rastall hydrostatic mass is compared with the observed baryonic mass; and (ii) the existence of DM, where the Rastall hydrostatic mass is employed to alleviate the hydrostatic mass bias.

In the comparison between the Rastall hydrostatic mass and the observed baryonic mass, using a sample of ten galaxy clusters, the best-fit result is obtained for $\lambda = 1.14\times10^{-1}$, yielding $\mathbf{M} = 1.07 \pm 0.11$. In this case, the unity value $\mathbf{M}=1$, which represents the ideal situation where the Rastall hydrostatic mass statistically matches the observed baryonic mass, lies within the uncertainty range and is also located in the high-likelihood region of the corresponding relative likelihood curve.

On the other hand, in the second scenario, by using \textcolor{black}{nine} galaxy clusters as the sample, we find that the best-fit value, \textcolor{black}{$\mathbf{M}=0.99\pm0.26$}, is obtained for \textcolor{black}{$\lambda=1.29\times10^{-3}$}. In this case, unity lies very close to the peak of the corresponding relative likelihood curve in the comparison between the Rastall hydrostatic mass and the observed lensing mass. \textcolor{black}{In the context of the scaling relation}, this result indicates that Rastall gravity can effectively alleviate the hydrostatic mass bias problem.

\textcolor{black}{We also calculate the goodness-of-fit for each case, using chi-squared $\chi^2$ and reduced chi-squared $\chi^2_\nu$ evaluations. The combination of the linear fit and the goodness-of-fit evaluations indicates that Rastall gravity provides a phenomenological framework that can improve certain aspects of the mass discrepancy problem, particularly at the level of scaling relations. However, it does not universally outperform other modified gravity models, such as non-local gravity, when evaluated using standard goodness-of-fit criteria.}

Nevertheless, a larger sample of galaxy clusters in the comparison between the Rastall hydrostatic mass and the gravitational lensing mass is required to further assess the viability of Rastall gravity and to constrain the Rastall parameter $\lambda$ using observational data.

\section*{Declaration of generative AI and AI-assisted technologies in the writing process}
During the preparation of this work, M. Lawrence Pattersons used ChatGPT (OpenAI) in order to support brainstorming and organization of some theoretical ideas, code development for numerical computation, and to improve the language and readability of the manuscript. All physical interpretations, model constructions, calculations, and final manuscript preparation were performed by the authors, who take full responsibility for the content. After using this tool, the authors reviewed and edited the content as needed and take full responsibility for the content of the paper.

\section*{Acknowledgements}
MLP is supported by the Indonesia Endowment Fund for Education Agency (LPDP). FA is supported by Physics Study Program, Faculty of Mathematics and Natural Science Education, Universitas Pendidikan Indonesia. We thank the Theoretical High Energy Physics Group at Institut Teknologi Bandung for discussions, insights, and hospitality. We would like to thank the anonymous reviewers for their thoughtful comments and efforts towards improving our manuscript.

\appendix
\setcounter{equation}{0}
\renewcommand{\theequation}{A\arabic{equation}}
\section{Statistical data of the linear fits}
The statistical data of the linear fits are shown by Table~\ref{appendixb}.
\begin{table*}
\centering
\caption{Linear fitting results between Rastall hydrostatic mass and observed baryonic and lensing masses for different values of the Rastall parameter $\lambda$.}
\label{tab:fit_combined}
\begin{tabular}{c c c c c c c}
\hline\hline
\multicolumn{7}{c}{\textbf{Rastall Hydrostatic Mass vs Observed Baryonic Mass}} \\
\hline
$\lambda$ 
& Intercept 
& Slope 
& RSS 
& Pearson $r$ 
& $R^2$ 
& Adj.\ $R^2$ \\
\hline
$0$ 
& $0.44 \pm 0.67$ 
& $8.24 \pm 0.86$ 
& $10.75$ 
& $0.96$ 
& $0.92$ 
& $0.91$ \\

$9.55 \times 10^{-2}$ 
& $0.05 \pm 0.07$ 
& $0.82 \pm 0.09$ 
& $0.11$ 
& $0.96$ 
& $0.92$ 
& $0.91$ \\

$1.14 \times 10^{-1}$ 
& $0.06 \pm 0.09$ 
& $1.07 \pm 0.11$ 
& $0.19$ 
& $0.96$ 
& $0.92$ 
& $0.91$ \\
\hline
\multicolumn{7}{c}{\textbf{Rastall Hydrostatic Mass vs Observed Lensing Mass}} \\
\hline
$0$ 
& \textcolor{black}{$0.48 \pm 1.74$} 
& \textcolor{black}{$0.88 \pm 0.23$} 
& \textcolor{black}{$34$} 
& \textcolor{black}{$0.82$} 
& \textcolor{black}{$0.67$} 
& \textcolor{black}{$0.62$} \\

\textcolor{black}{$1.14 \times 10^{-3}$} 
& \textcolor{black}{$0.55 \pm 1.92$} 
& \textcolor{black}{$0.98 \pm 0.26$} 
& \textcolor{black}{$42.71$} 
& \textcolor{black}{$0.82$} 
& \textcolor{black}{$0.67$} 
& \textcolor{black}{$0.62$} \\

\textcolor{black}{$1.29 \times 10^{-3}$} 
& \textcolor{black}{$0.56 \pm 1.92$} 
& \textcolor{black}{$0.99 \pm 0.26$} 
& \textcolor{black}{$44.11$} 
& \textcolor{black}{$0.82$} 
& \textcolor{black}{$0.67$} 
& \textcolor{black}{$0.62$} \\
\hline\hline
\end{tabular}
\label{appendixb}
\end{table*}
%




\end{document}